\pdfoutput=1

\documentclass[11pt,twoside,a4paper,cmspaper,final,collab]{cms-tdr}
\def\svnVersion{daf9521-D}\def\svnDate{2020/09/25}\def\cmsCernNoTag{CERN-EP-2020-144}\def\cmsCernDate{\today}\def\cmsMessage{"Published in Physical Review D as \href{http://dx.doi.org/10.1103/PhysRevD.102.092001}{\doi{10.1103/PhysRevD.102.092001}.}"}

\begin{document}\cmsNoteHeader{SMP-18-004}

\hyphenation{had-ron-i-za-tion}
\hyphenation{cal-or-i-me-ter}
\hyphenation{de-vices}
\newlength\cmsTabSkip\setlength{\cmsTabSkip}{1ex}
\newlength\cmsFigWidthTwo
\ifthenelse{\boolean{cms@external}}{\setlength\cmsFigWidthTwo{0.23\textwidth}}{\setlength\cmsFigWidthTwo{0.49\textwidth}}
\newlength\cmsFigWidthFix
\ifthenelse{\boolean{cms@external}}{\setlength\cmsFigWidthFix{0.23\textwidth}}{\setlength\cmsFigWidthFix{0.44\textwidth}}
\ifthenelse{\boolean{cms@external}}{\providecommand{\cmsLeft}{upper\xspace}}{\providecommand{\cmsLeft}{left\xspace}}
\ifthenelse{\boolean{cms@external}}{\providecommand{\cmsRight}{lower\xspace}}{\providecommand{\cmsRight}{right\xspace}}
\providecommand{\cmsTable}[1]{\resizebox{\textwidth}{!}{#1}}
\providecommand{\experr}{\ensuremath{\,\text{(exp)}}\xspace}

\newcommand{\Etmissproj}{\ensuremath{\pt^{\text{miss,proj}}}}
\newcommand{\Mettrackproj}{\ensuremath{\pt^{\text{miss,track~proj}}}}
\newcommand{\mll}{\ensuremath{m_{\Pell\Pell}}\xspace}
\newcommand{\ptmax}{\ensuremath{\pt^{\Pell\text{\,max}}}\xspace}
\newcommand{\ptmin}{\ensuremath{\pt^{\Pell\text{\,min}}}\xspace}
\newcommand{\delphill}{\ensuremath{\Delta\phi_{\Pell\Pell}}\xspace}
\newcommand{\WW}{\ensuremath{\PW\PW}\xspace}
\newcommand{\WWpm}{\ensuremath{\PWp\PWm}\xspace}
\newcommand{\WWpt}{\ensuremath{\pt^{\PW\PW}}\xspace}
\newcommand{\memu}{\ensuremath{m_{\Pe\PGm}}\xspace}
\newcommand{\qT}{\pt^{\Pell\Pell}}
\newcommand{\ptj}{\pt^{\mathrm{J}}\xspace}
\newcommand{\etaj}{\ensuremath{\eta^{\mathrm{J}}\xspace}}
\newcommand{\BFWlept}{{\mathcal{B}}(\PW\to\Pell\nu)}
\newcommand{\sigmatotTA}{\sigma_{\text{SC}}^{\text{tot}}}
\newcommand{\sigmatotRF}{\sigma_{\text{RF}}^{\text{tot}}}
\newcommand{\sigmafidTA}{\sigma^{\text{fid}}}
\newcommand{\sigmaNNLO}{\sigma_{\text{tot}}^{\text{NNLO}}}
\newcommand{\sigmaNNLOval}{118.8\pm3.6\unit{pb}}
\newcommand{\theLumi}{35.9\pm0.9\fbinv}
\newcommand{\pb}{\unit{pb}}
\newcommand{\SDY} {S_{\text{DY}}}
\newcommand{\STT} {S_{\ttbar}}
\newcommand{\SDYmin} {S_{\text{DY}}^{\text{min}}}
\newcommand{\STTmin} {S_{\ttbar}^{\text{min}}}
\newcommand{\NJ}{\ensuremath{N_{\mathrm{J}}}\xspace}
\newcommand{\RPU}{{\mathbf{R}}_{\text{PU}}}
\newcommand{\RJER}{{\mathbf{R}}_{\text{det}}}
\newcommand{\RPUI}{{{\mathbf{R}}_{\text{PU}}}^{-1}}
\newcommand{\RJERI}{{{\mathbf{R}}_{\text{DET}}}^{-1}}
\newcommand{\CSVvv} {{\textsc{csv}\,v2}\xspace}
\newcommand{\Niexp} {{N_i^{\text{exp}}}}
\newcommand{\Niobs} {{N_i^{\text{obs}}}}

\cmsNoteHeader{SMP-18-004} 
\title{\texorpdfstring{$\WWpm$}{WW} boson pair production in proton-proton collisions at \texorpdfstring{$\sqrt{s} = 13\TeV$}{sqrt(s) = 13 TeV}}

\date{\today}

\abstract{A measurement of the \WWpm boson pair production cross section in proton-proton collisions at $\sqrt{s} = 13\TeV$ is presented. The data used in this study are collected with the CMS detector at the CERN LHC and correspond to an integrated luminosity of 35.9\fbinv. The \WWpm candidate events are selected by requiring two oppositely charged leptons (electrons or muons). Two methods for reducing background contributions are employed.  In the first one, a sequence of requirements on kinematic quantities is applied allowing a measurement of the total production cross section: $117.6 \pm 6.8\pb$, which agrees well with the theoretical prediction. Fiducial cross sections are also reported for events with zero or one jet, and the change in the zero-jet fiducial cross section with the jet transverse momentum threshold is measured.  Normalized differential cross sections are reported within the fiducial region.  A second method for suppressing background contributions employs two random forest classifiers.  The analysis based on this method includes a measurement of the total production cross section and also a measurement of the normalized jet multiplicity distribution in \WWpm events. Finally, a dilepton invariant mass distribution is used to probe for physics beyond the standard model in the context of an effective field theory, and constraints on the presence of dimension-6 operators are derived.}

\hypersetup{%
pdfauthor={CMS Collaboration},%
pdftitle={WW boson pair production in proton-proton collisions at sqrt(s)=13 TeV},%
pdfsubject={CMS},%
pdfkeywords={CMS, physics, W boson pairs}}

\maketitle

\section{Introduction}\label{sec:intro}
The standard model (SM) description of electroweak and strong interactions can 
be tested through measurements of the \WWpm boson pair production cross section at 
a hadron collider. Aside from tests of the SM, \WWpm production represents 
an important background for new particle searches. The \WWpm cross section
has been measured in proton-antiproton collisions at
$\sqrt{s} = 1.96\TeV$~\cite{Abazov:2009ys,Aaltonen:2009aa}
and in proton-proton ($\Pp\Pp$) collisions at 7 and 
8\TeV~\cite{ATLAS:2012mec, Chatrchyan:2013yaa, Khachatryan:2015sga, Aad:2016wpd}. 
More recently, the ATLAS Collaboration published measurements with $\Pp\Pp$ collision data at
13\TeV~\cite{ATLAS:2019nkz}.

The SM production of \WWpm pairs proceeds mainly through three processes:
the dominant $\cPq\cPaq$ annihilation process;
the $\Pg\Pg\to\PWp\PWm$ process, which occurs at higher order
in perturbative quantum chromodynamics (QCD); and the Higgs boson process $\PH\to\WWpm$,
which is roughly ten times smaller than the other processes and 
is considered a background in this analysis.
A calculation of the \WWpm production cross section in $\Pp\Pp$ collisions at
$\sqrt{s} = 13\TeV$ gives the value $118.7^{+3.0}_{-2.6}\pb$~\cite{Gehrmann:2014fva}.
This calculation includes the $\cPq\cPaq$ annihilation process calculated at
next-to-next-to-leading order (NNLO) precision in perturbative QCD and a
contribution of 4.0\pb from the $\Pg\Pg\to\PWp\PWm$ gluon fusion process
calculated at leading order (LO). The uncertainties reflect the dependence of the
calculation on the QCD factorization and renormalization scales.
For the analysis presented in this paper,
the $\Pg\Pg\to\WWpm$ contribution is corrected by a factor of 1.4, which comes from the ratio
of the $\Pg\Pg\to\WWpm$ cross section at next-to-leading order (NLO) to the
same cross section at LO~\cite{Caola:2015rqy}.  A further adjustment of $-1.2$\% for
the  $\cPq\cPaq$ annihilation process is applied to account for electroweak corrections~\cite{WWEWKCorr}.
Our evaluation of uncertainties from parton distribution functions (PDFs) and 
the strong coupling $\alpS$ amounts to 2.0\pb. Taking all corrections
and uncertainties together, the theoretical cross section used in this paper for the inclusive
\WWpm production at $\sqrt{s} = 13\TeV$ is $\sigmaNNLO = \sigmaNNLOval$.

This paper reports studies of \WWpm production in $\Pp\Pp$ collisions at $\sqrt{s} = 13\TeV$ with the CMS detector at the CERN LHC.
Two analyses are performed using events that contain a pair of oppositely charged leptons (electrons or muons);
they differ in the way background contributions are reduced.
The first method is based on techniques described in
Refs.~\cite{Aad:2016wpd, Chatrchyan:2013yaa, Khachatryan:2015sga};
the analysis based on this method is referred to as the ``sequential cut analysis.''
A second, newer approach makes use of random forest classifiers~\cite{Breiman,Ho1998,Caruana2005}
trained with simulated data to differentiate signal events from Drell--Yan (DY)
and top quark backgrounds; this analysis is referred to as the ``random forest analysis.''

The two methods complement one another.  The sequential cut analysis 
separates events with same-flavor (SF) or different-flavor (DF) lepton pairs, and also
events with zero or one jet.  As a consequence, background contributions
from the Drell--Yan production of lepton pairs can be controlled.  Furthermore,
the impact of theoretical uncertainties due to missing higher-order QCD calculations is kept
under control through access to both the zero- and one-jet final states.
The random forest analysis does not separate SF and DF
lepton pairs and does not separate events with different jet multiplicities.
Instead, it combines kinematic and topological quantities to achieve
a high sample purity.  The contamination from top quark events,
which is not negligible in the sequential cut analysis, is significantly smaller in the random forest analysis.
The random forest technique allows for flexible control over the
top quark background contamination, which is exploited to study the jet
multiplicity in \WWpm signal events.  However, the sensitivity of the random forest
to QCD uncertainties is significantly larger than that of
the sequential cut analysis, as discussed in Section~\ref{sec:TotalCrossSection}.

Total \WWpm production cross sections are reported in Section~\ref{sec:TotalCrossSection}
for both analyses based on fits to the observed yields.
Cross sections in a specific fiducial region are reported in Section~\ref{sec:fiducialX}
for the sequential cut analysis;
these cross sections are separately reported for $\WWpm\to\Pell^+\nu\Pell^-\Pagn$
events with zero or one jet (\Pell~refers to electrons and muons).
Also, the change in the zero-jet \WWpm cross section with variations in the jet
transverse momentum (\pt) threshold is measured.

Normalized differential cross sections within the fiducial region are also reported in Section~\ref{sec:diffWWxsec}.
The normalization reduces both theoretical and experimental uncertainties.  The impact
of experimental resolutions is removed using a fitting technique that builds templates
of reconstructed quantities mapped onto generator-level quantities.
Comparisons to NLO predictions are presented.

The distribution of exclusive jet multiplicities for \WWpm production is
interesting given the sensitivity of previous results to a ``jet veto'' in which
events with one or more jets were rejected~\cite{ATLAS:2012mec,Aad:2016wpd,Chatrchyan:2013yaa,Aaltonen:2009aa}.
In Section~\ref{sec:NJets}, this paper reports a measurement of the normalized jet multiplicity distribution
based on the random forest analysis.

Finally, the possibility of anomalous production of \WWpm events that can be modeled by
higher-dimensional operators beyond the dimension-4 operators
of the SM is probed using events with an electron-muon final state.
Such operators arise in an effective field theory expansion of the Lagrangian
and each appears with its own Wilson coefficient~\cite{Weinberg:1978kz,Degrande:2012wf}.
Distributions of the electron-muon
invariant mass \memu are used because they are robust against mismodeling of the \WWpm transverse
boost, and are sensitive to the value of the Wilson coefficients associated with the dimension-6 operators.
The observed distributions provide no evidence for anomalous events. 
Limits are placed on the coefficients associated with dimension-6 operators in Section~\ref{sec:aTGC}.

\section{The CMS detector}\label{sec:detector}
The central feature of the CMS apparatus is a superconducting solenoid of 
6\unit{m} internal diameter, providing a magnetic field of 3.8\unit{T}. 
Within the solenoid volume are a silicon pixel and strip tracker, a lead 
tungstate crystal electromagnetic calorimeter (ECAL), and a brass and 
scintillator hadron calorimeter (HCAL), each composed of a barrel and 
two endcap sections. Forward calorimeters extend the pseudorapidity ($\eta$) coverage 
provided by the barrel and endcap detectors. Muons are detected in 
gas-ionization chambers embedded in the steel flux-return yoke outside the solenoid. 
The first level of the CMS trigger system~\cite{CMSTrigger}, composed of custom hardware processors, 
is designed to select the most interesting events within a time interval less than 4\rm{\mus}, 
using information from the calorimeters and muon detectors,  with the 
output rate of up to 100\unit{kHz}. The high-level trigger processor farm 
further reduces the event rate to about 1\unit{kHz} before data storage.
A more detailed description of the CMS detector, together with a definition of the
coordinate system used and the relevant kinematic variables, can be found 
in Ref.~\cite{Chatrchyan:2008zzk}.

\section{Data and simulated samples}\label{sec:samples}
A sample of $\Pp\Pp$ collision data collected in 2016 with the CMS experiment 
at the LHC at $\sqrt{s} = 13\TeV$ is used for this analysis;
the total integrated luminosity is $\theLumi$.

Events are stored for analysis if they satisfy the selection criteria of
online triggers~\cite{CMSTrigger} requiring the presence of one 
or two isolated leptons (electrons or muons) with high \pt. 
The lowest \pt thresholds for the double-lepton triggers are $17\GeV$ for the
leading lepton and 12 (8)\GeV  when the trailing lepton is
an electron (muon).  The single-lepton triggers have \pt thresholds 
of 25 and 20\GeV for electrons and muons, respectively. 
The trigger efficiency is measured using $\PZ\to\Pell^+\Pell^-$ events
and is larger than 98\% for \WWpm events with an uncertainty of about~1\%.

Several Monte Carlo (MC) event generators are used to simulate the signal and 
background processes. The simulated samples are used to optimize the event 
selection, evaluate selection efficiencies and systematic uncertainties, and 
compute expected yields. The production of \WWpm events via $\Pq\Paq$ annihilation 
($\Pq\Paq\to\WWpm$) is generated at NLO precision 
with \POWHEG{\sc\,v2}~\cite{Nason:2004rx,Frixione:2007vw,powheg:2010,Nason:2013ydw,Alioli:2008gx,Alioli:2008tz},
and \WWpm production via gluon fusion ($\Pg\Pg\to\WWpm$) is generated 
at LO using \MCFM~v7.0~\cite{Campbell:2010ff}. 
The production of Higgs bosons is generated with \POWHEG~\cite{Alioli:2008tz} 
and $\PH\to\WWpm$ decays are generated with {\sc JHUGen v5.2.5}~\cite{jhugen}.
Events for other diboson and triboson production processes are generated at NLO precision with 
{\MGvATNLO 2.2.2}~\cite{Alwall:2014hca}. 
The same generator is used for simulating \cPZ{+}jets, which includes Drell--Yan production, and $\PW\gamma^\ast$ event samples.
Finally, the top quark final states \ttbar and $\cPqt\PW$ are 
generated at NLO precision with \POWHEG~\cite{Campbell:2014kua,Re:2010bp}.
The \PYTHIA~8.212~\cite{Sjostrand:2015} package with the CUETP8M1 parameter set (tune)~\cite{CMS:2014sva}
and the  NNPDF~2.3~\cite{nnpdf23} PDF set
are used for hadronization, parton showering, and the underlying event simulation.
For top quark processes, the  NNPDF~3.0 PDF set~\cite{nnpdf} and
the CUETP8M2T4 tune~\cite{CMS-PAS-TOP-16-021} are used.

The quality of the signal modeling is improved by applying weights to
the \WWpm \POWHEG events such that the NNLO calculation~\cite{Gehrmann:2014fva}
of transverse momentum spectrum of the \WWpm system, \WWpt,
is reproduced. 

For all processes, the detector response is simulated using a detailed 
description of the CMS detector, based on the \GEANTfour
package~\cite{Agostinelli:2002hh}. Events are reconstructed with 
the same algorithms as for data. The simulated samples 
include additional interactions per bunch crossing (pileup) with 
a vertex multiplicity distribution that closely matches the observed one.

\section{Event reconstruction}\label{sec:objects}
Events are reconstructed using the CMS particle-flow (PF) algorithm~\cite{pflow}, 
which combines information from the tracker, calorimeters, and muon systems to
create objects called PF candidates that are subsequently identified as
charged and neutral hadrons, photons, muons, and electrons. 

The primary $\Pp\Pp$ interaction vertex is defined to be the one with the largest
value of the sum of $\pt^2$ for all physics objects associated with that vertex.
These objects include jets clustered using the jet finding algorithm~\cite{antikt,Cacciari:2011ma} 
with the tracks assigned to the primary vertex as inputs, and the associated missing transverse momentum vector. 
All neutral PF candidates and charged PF candidates 
associated with the primary vertex are clustered into jets using the anti-$\kt$ 
clustering algorithm~\cite{antikt} with a distance parameter of  $R=0.4$.
The transverse momentum imbalance $\ptvecmiss$ is the negative vector sum of the transverse momenta of all
charged and neutral PF candidates; its magnitude is denoted by \ptmiss.
The effects of pileup are mitigated as described in Ref.~\cite{CMS-PAS-JME-10-003,Sirunyan:2020foa}.

Jets originating from \PQb quarks are identified by a multivariate algorithm called
the combined secondary vertex algorithm \CSVvv~\cite{Chatrchyan:2012jua,Sirunyan:2017ezt},
which combines information from tracks, secondary vertices, and low-momentum 
electrons and muons associated with the jet.
Two working points are used in this analysis for jets with $\pt>20\GeV$.
The ``loose'' working point has an efficiency of approximately 88\% for jets originating from the 
hadronization of \PQb quarks typical in \ttbar events and a mistag rate of about 10\%
for jets originating from the hadronization of light-flavor quarks or gluons.
The ``medium'' working point has a \PQb~tagging efficiency of about 64\%
for \PQb~jets in \ttbar  events and
a mistag rate of about 1\% for light-flavor quark and gluon jets.

Electron candidates are reconstructed from clusters in the ECAL
that are matched to a track reconstructed with a Gaussian-sum 
filter algorithm~\cite{Khachatryan:2015hwa}. The track is required 
to be consistent with originating from the primary vertex. The sum of the 
\pt of PF candidates within a cone of size 
$\Delta R = \sqrt{\smash[b]{(\Delta\eta)^2+(\Delta\phi)^2}} < 0.3$ around the electron 
direction, excluding the electron itself, is required to be less than about 6\% of the electron \pt.
Charged PF candidates are included in the isolation sum only if 
they are associated with the primary vertex.  The average contribution from 
neutral PF candidates not associated with the primary vertex, estimated 
from simulation as a function of the energy density in the event and the 
$\eta$ direction of the electron candidate, is subtracted before comparing 
to the electron momentum.

Muon candidates are reconstructed by combining signals from the muon 
subsystems together with those from the tracker~\cite{Chatrchyan:2013sba,Sirunyan:2018fpa}. 
The track reconstructed in the silicon pixel and strip detector  must be consistent with originating from the primary 
vertex. The sum of the \pt of the additional PF candidates 
within a cone of size $\Delta R < 0.4$ around the muon direction
is required to be less than 15\% of the muon \pt
after applying a correction for neutral PF candidates not associated with
the primary vertex, analogous to the electron case.

\section{Event selection}\label{sec:selection}
The key feature of the \WWpm channel is the presence of two oppositely charged
leptons that are isolated from any jet activity and have relatively large \pt.
The two methods for isolating a \WWpm signal, the sequential cut method
and the random forest method, both require two oppositely charged,
isolated electrons or muons that have sufficient \pt to ensure good
trigger efficiency.   The lepton reconstruction, selection, and isolation criteria
are the same for the two methods as are most of the kinematic requirements
detailed below.

The largest background contributions come from the Drell--Yan production of lepton pairs
and \ttbar events in which both top quarks decay leptonically.  Drell--Yan events
can be suppressed by selecting events with one electron and one muon (\ie, DF leptons)
and by applying a veto of the \cPZ boson resonant peak in events with
SF leptons.  Contributions from \ttbar events can be reduced by rejecting
events with \PQb-tagged jets.

Another important background contribution arises from events with one
or more jets produced in association with a single \PW boson.  A nonprompt
lepton from a jet could be selected with charge opposite to that of the
prompt lepton from the \PW boson decay.  This background contribution is estimated
with two techniques based on specially selected events.  In the sequential cut
analysis, the calculation hinges on the probability for a nonprompt lepton to
be selected, whereas in the random forest selection, it depends on a sample of events
with two leptons of equal charge.

Except where noted, \WWpm events with $\tau$ leptons decaying to electrons or
muons are included as signal.

\subsection{Sequential cut selection}
\label{sec:selectionTA}

The sequential cut selection imposes a set of discrete requirements
on kinematic and topological quantities and on a multivariate analysis tool 
to suppress Drell--Yan background in events with SF leptons.

The lepton \pt requirements ensure a good reconstruction and identification efficiency:
the leading lepton must have $\ptmax>25\GeV$, and the trailing lepton
must have $\ptmin>20\GeV$. Pseudorapidity
ranges are designed to cover regions of good reconstruction quality: for electrons,
the ECAL supercluster must satisfy $\abs{\eta} < 1.479$ or $1.566 < \abs{\eta} < 2.5$ and
for muons, $\abs{\eta} < 2.4$.  To avoid low-mass resonances and leptons
from decays of hadrons, the dilepton invariant mass must be large enough:
$\mll>20\GeV$.  The transverse momentum of the lepton pair
is required to satisfy $\qT>30\GeV$ to reduce background
contributions from nonprompt leptons.  Events with a third, loosely identified
lepton with $\pt>10\GeV$ are rejected to reduce background contributions
from $\PW\PZ$ and $\PZ\PZ$ (\ie, $\PV\PZ$) production.

The missing transverse momentum is required 
to be $>$20\GeV. In order to make the analysis insensitive to  
instrumental \ptmiss caused by mismeasurements of the lepton momenta,
a so-called ``projected $\ptmiss$'', denoted $\Etmissproj$, is defined as follows.
The lepton closest to the $\ptvecmiss$ vector is identified and the azimuthal
angle $\Delta\phi$ between the $\ptvec$ of the lepton and $\ptvecmiss$ is computed.
The quantity $\Etmissproj$ is the perpendicular component of $\ptvecmiss$ with respect
to $\ptvec$.  When $\abs{\Delta\phi} < \pi/2$, $\Etmissproj$ is
required to be larger than 20\GeV.
The same requirement is imposed using the projected $\ptvecmiss$ vector reconstructed from 
only the charged PF candidates associated with the primary vertex:
$\Mettrackproj>20\GeV$.

The selection criteria are tightened for SF final states where the contamination 
from Drell--Yan events is much larger. Events with $\mll$ within 15\GeV
of the $\PZ$~boson mass $m_{\PZ}$ are discarded, and the minimum $\mll$ is
increased to 40\GeV.  The \ptmiss requirement is raised
to 55\GeV.  Finally, a multivariate classifier called DYMVA~\cite{tmva,HiggsWW2016} 
based on a boosted decision tree is used to discriminate against the
Drell--Yan background.

Only events with zero or one reconstructed jet with $\ptj>30\GeV$ and 
$\abs{\etaj} < 4.7$ are used in the analysis.  Jets falling within
$\Delta R < 0.4$ of a selected lepton are discarded.
To suppress top quark background contributions, events with one or more jets tagged as \PQb~jets
using the \CSVvv loose working point and with $\pt^{\PQb}>20\GeV$ are also  rejected.

\begin{figure}
\centering
\includegraphics[width=\cmsFigWidthFix]{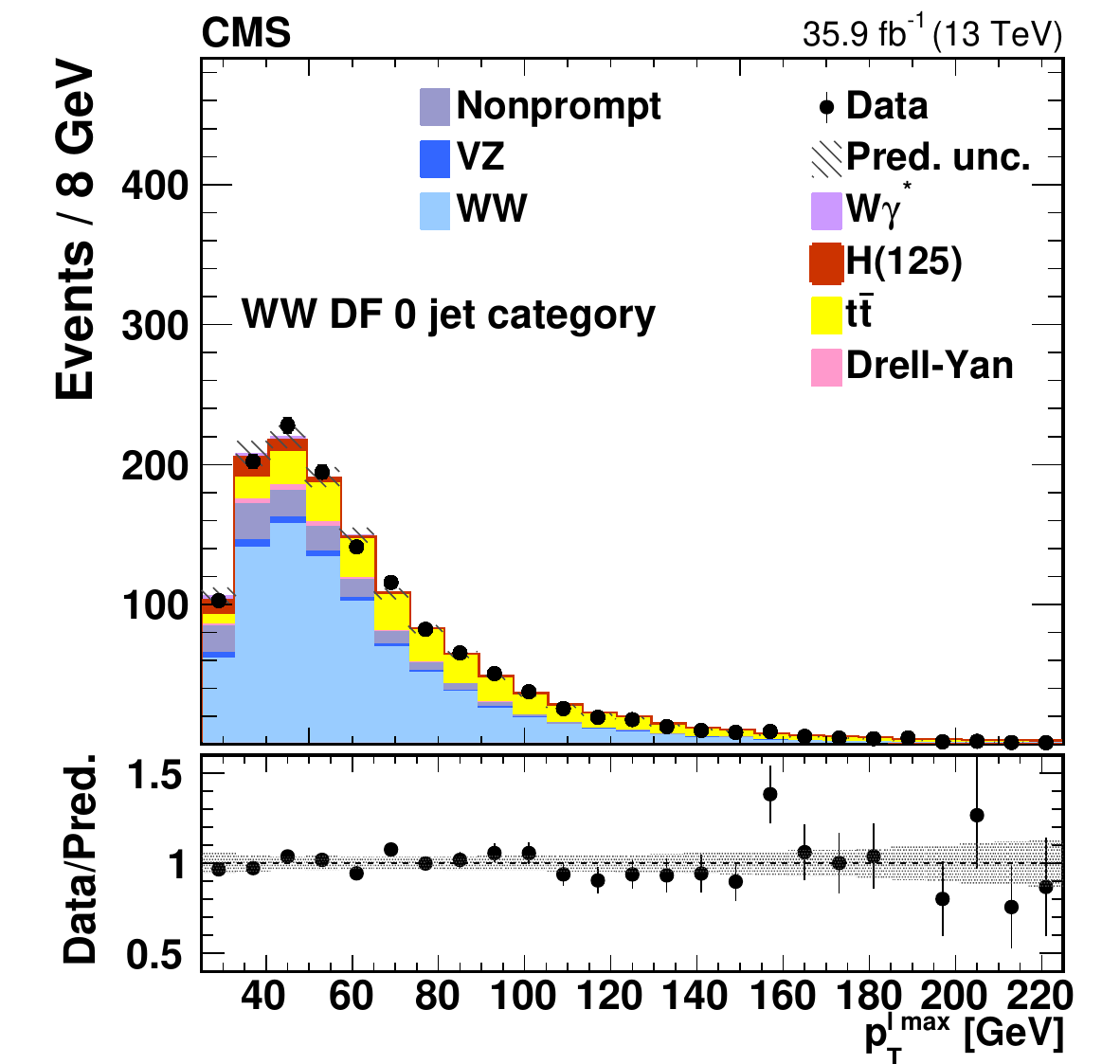}
\includegraphics[width=\cmsFigWidthFix]{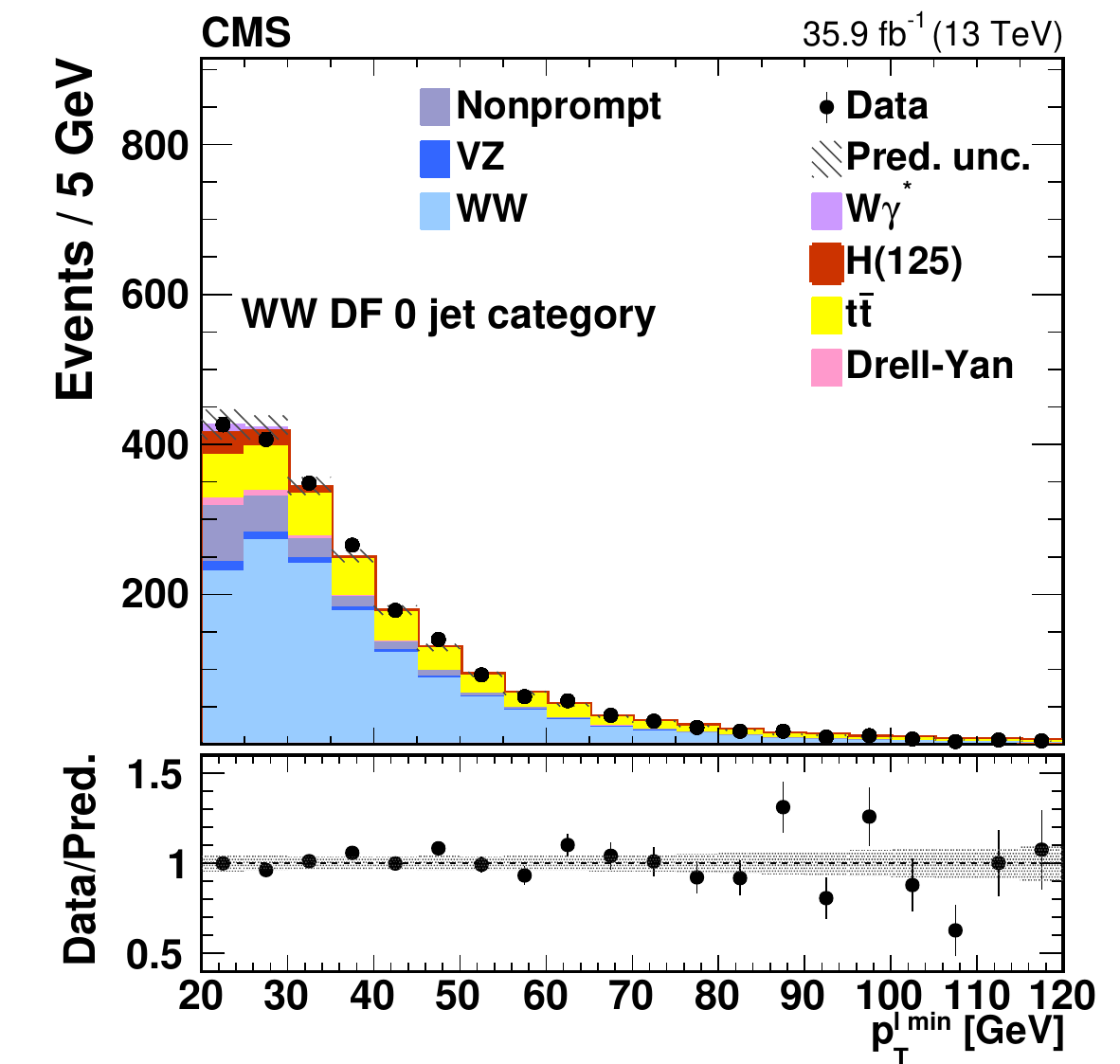}\\
\includegraphics[width=\cmsFigWidthFix]{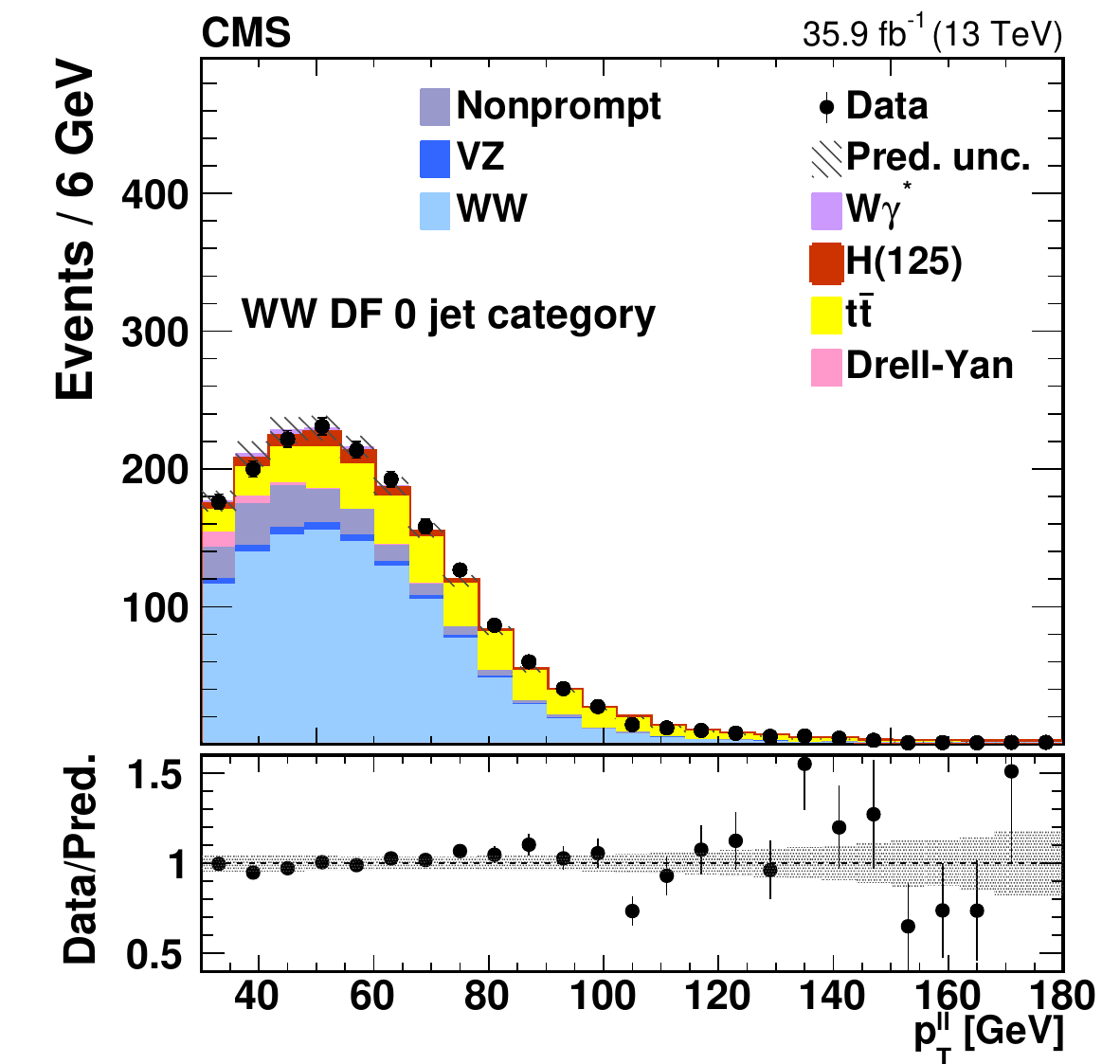}
\includegraphics[width=\cmsFigWidthFix]{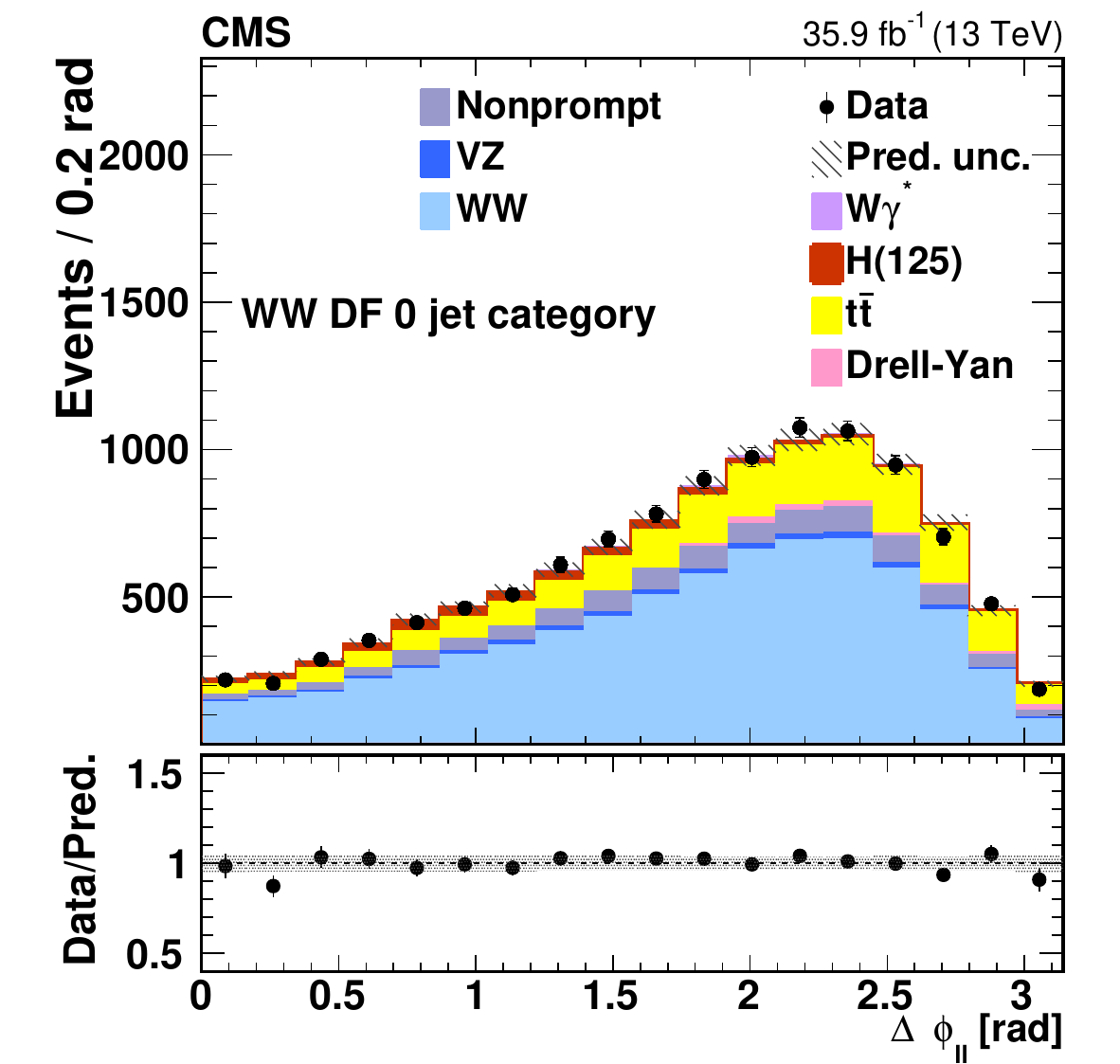}\\
\includegraphics[width=\cmsFigWidthFix]{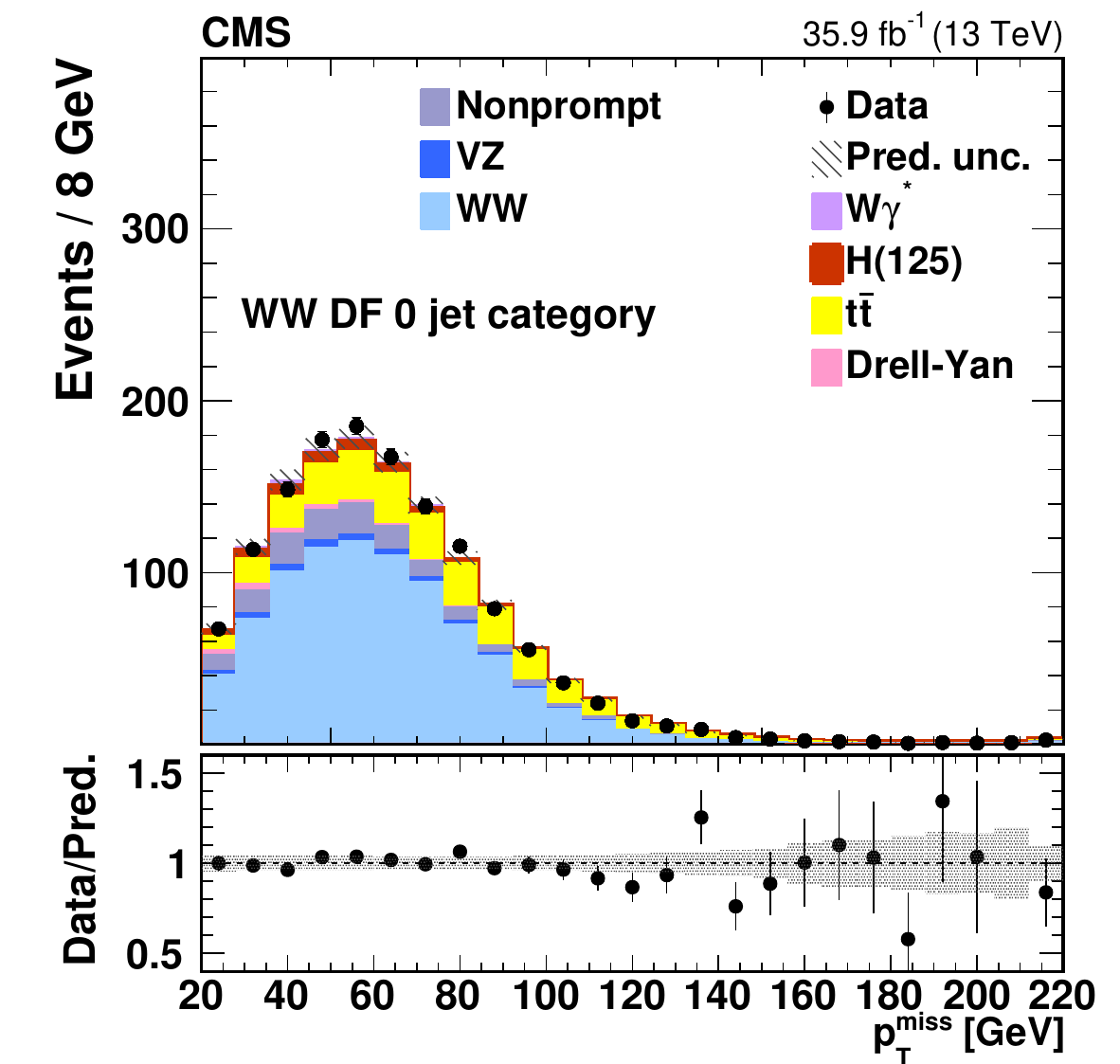}
\includegraphics[width=\cmsFigWidthFix]{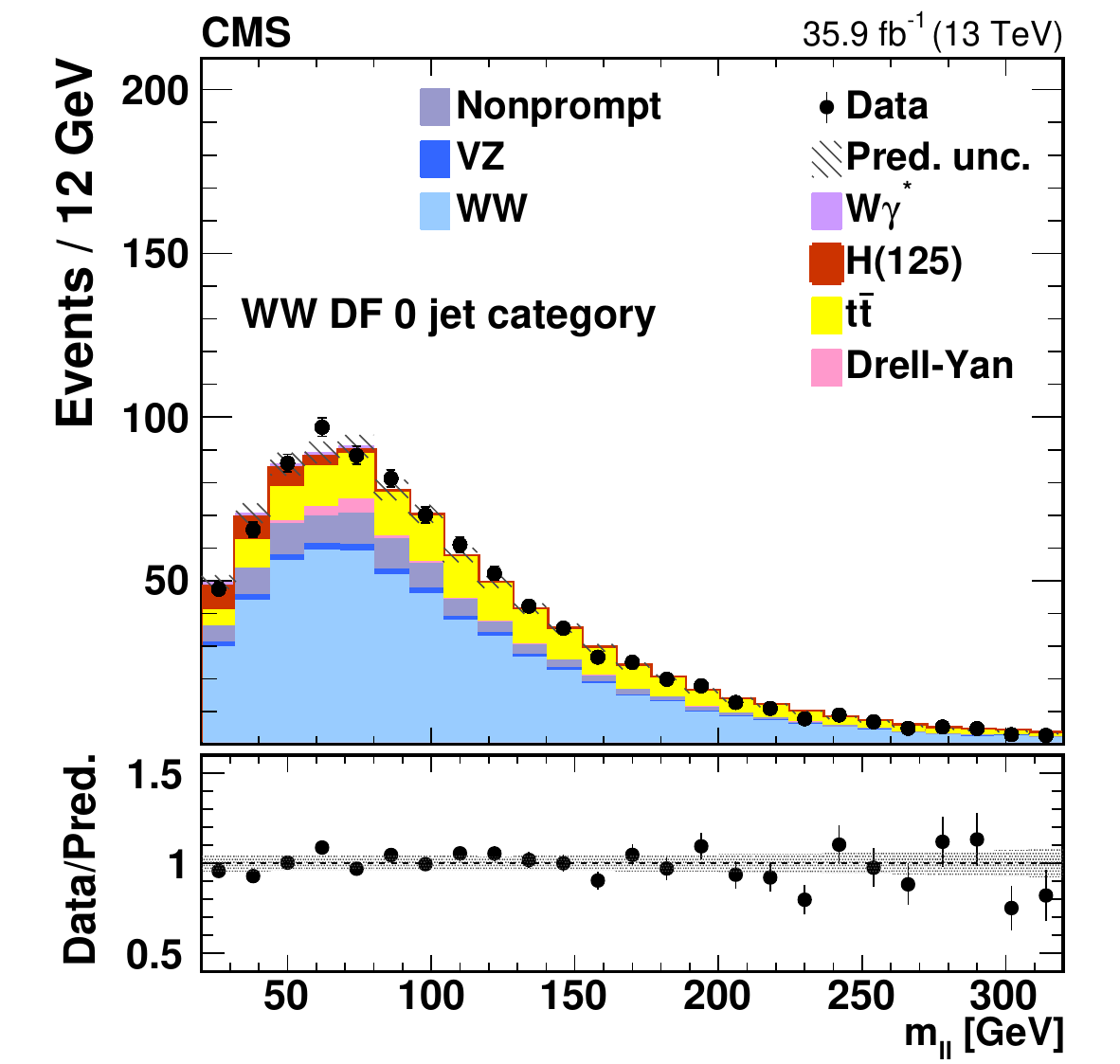}
\caption[.]{\label{fig:trad0jet}
Kinematic distributions for events with zero jets and DF leptons
in the sequential cut analysis.
The distributions show the leading and trailing lepton \pt ($\ptmax$ and $\ptmin$),
the dilepton transverse momentum  $\qT$, the azimuthal angle between the two leptons $\delphill$,
the missing transverse momentum \ptmiss, and the dilepton invariant mass $\mll$.
The error bars on the data points represent the statistical uncertainty of the data, and
the hatched areas represent the combined systematic and statistical uncertainty
of the predicted yield in each bin.  The last bin includes the overflow.}
\end{figure}

\begin{figure}
\centering
\includegraphics[width=\cmsFigWidthTwo]{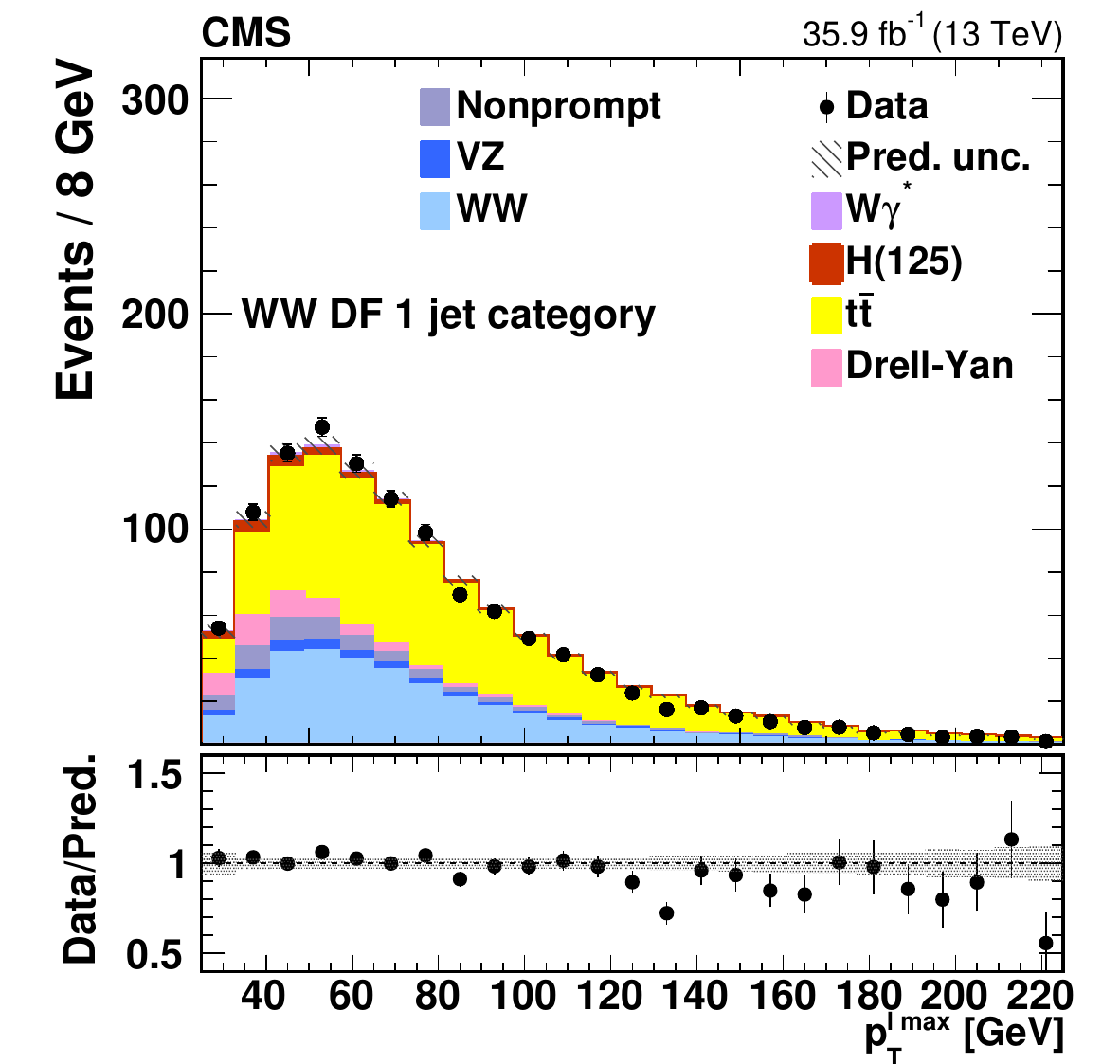}
\includegraphics[width=\cmsFigWidthTwo]{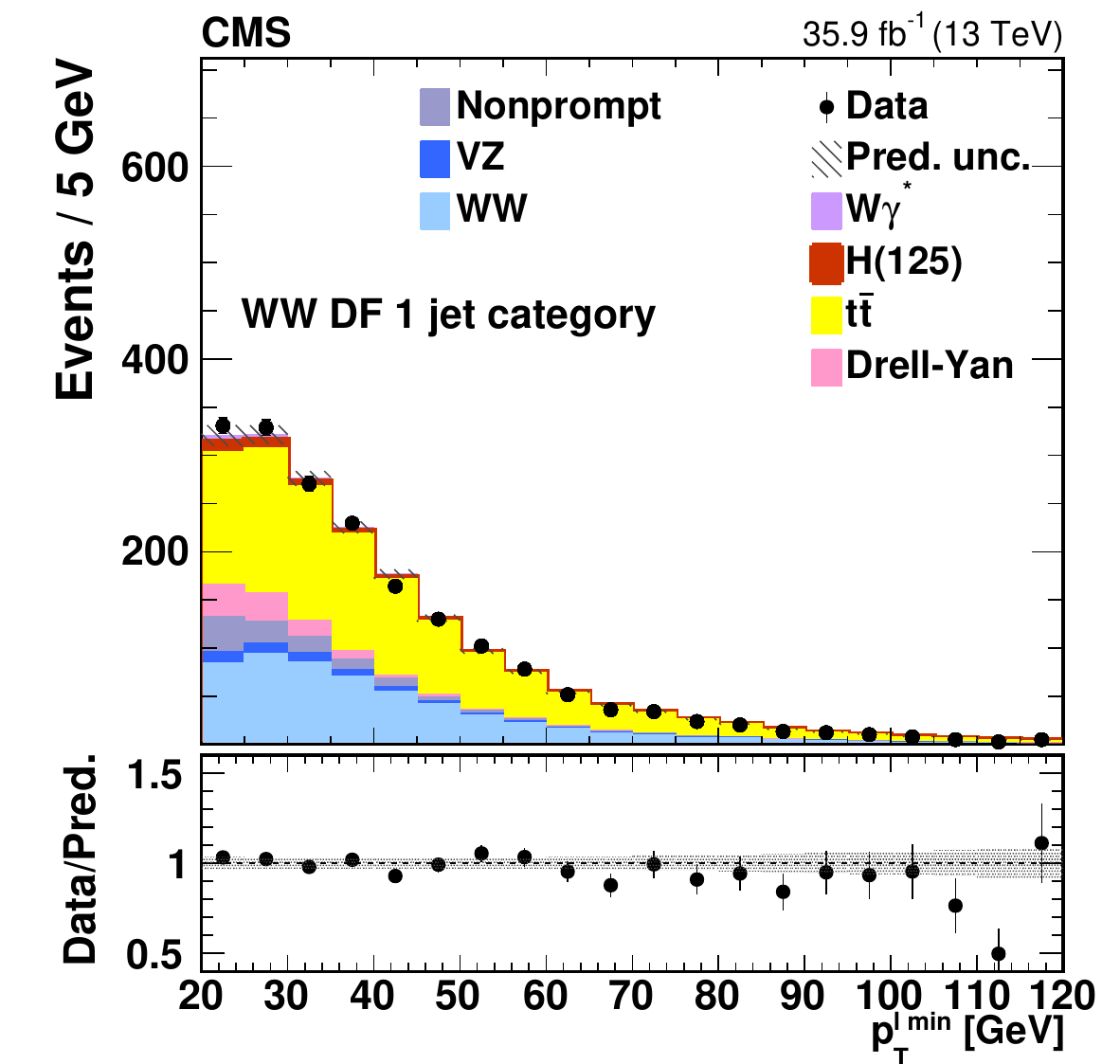}\\
\includegraphics[width=\cmsFigWidthTwo]{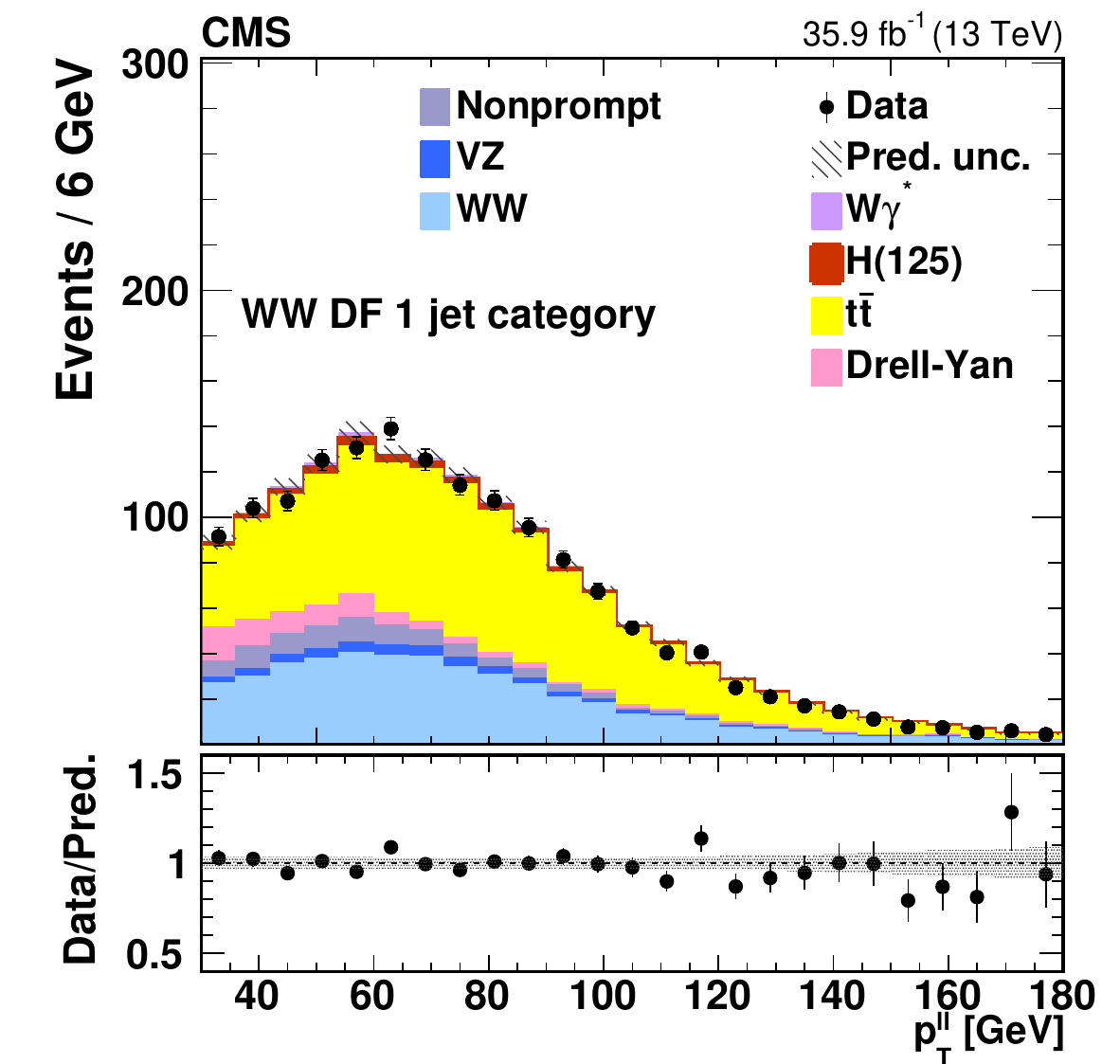}
\includegraphics[width=\cmsFigWidthTwo]{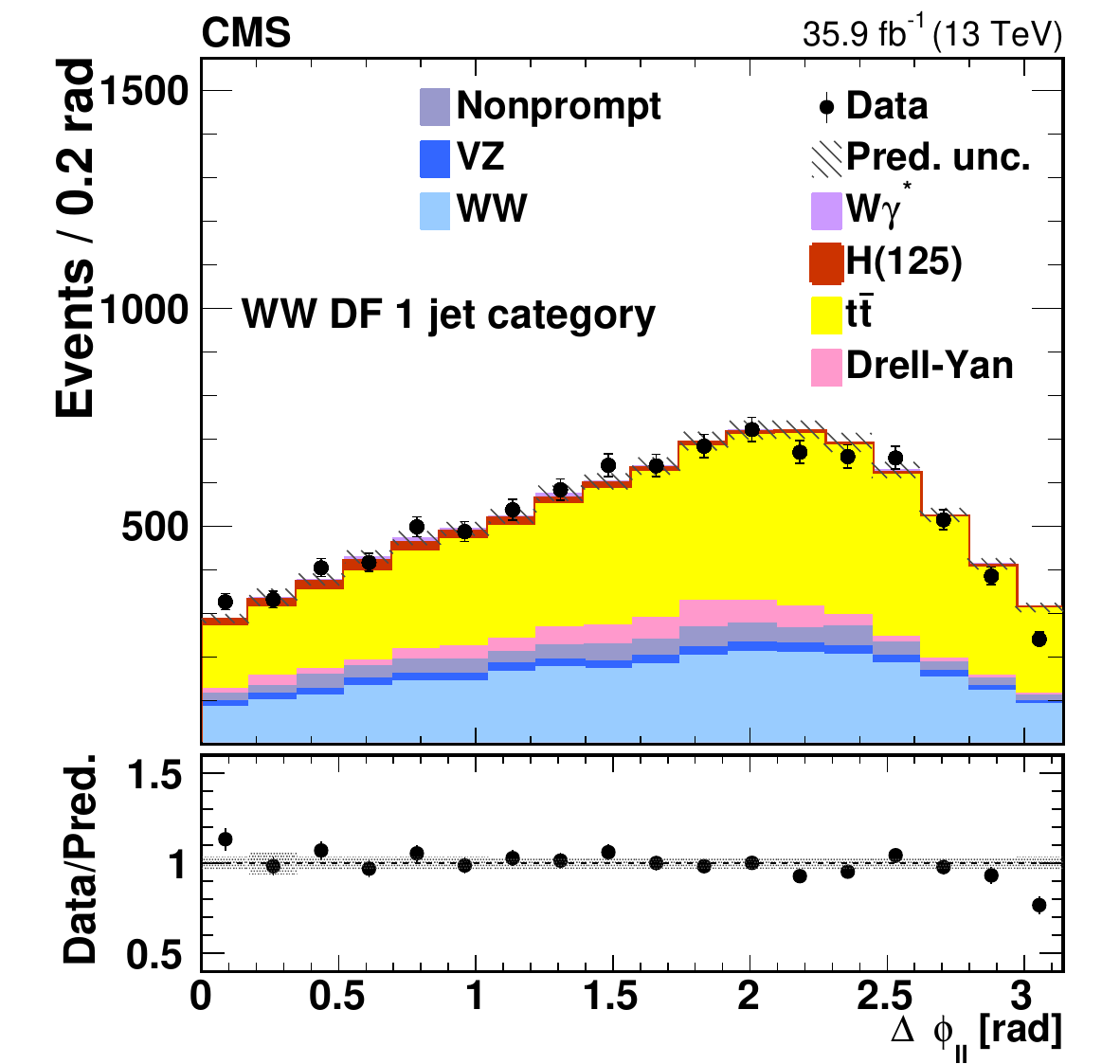}\\
\includegraphics[width=\cmsFigWidthTwo]{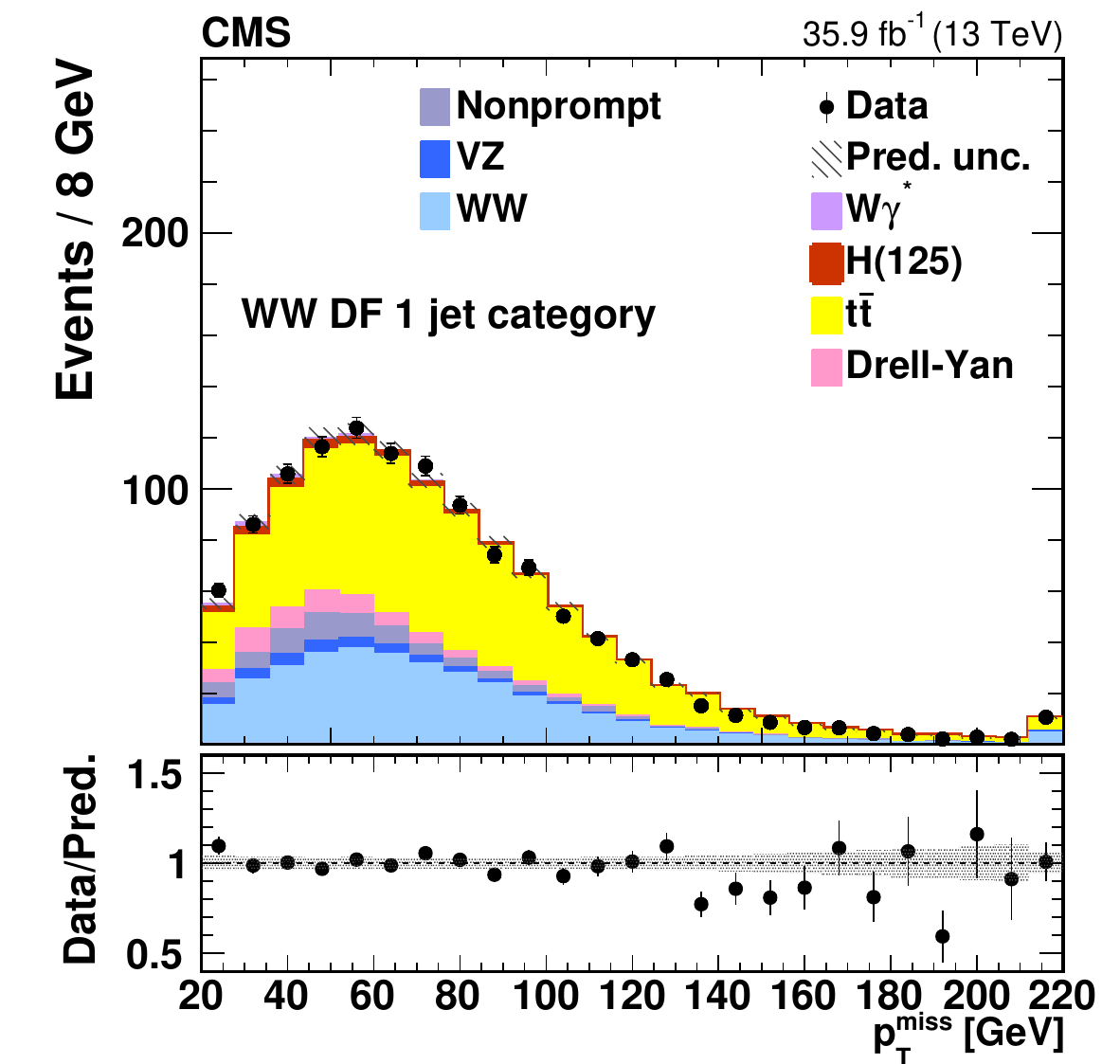}
\includegraphics[width=\cmsFigWidthTwo]{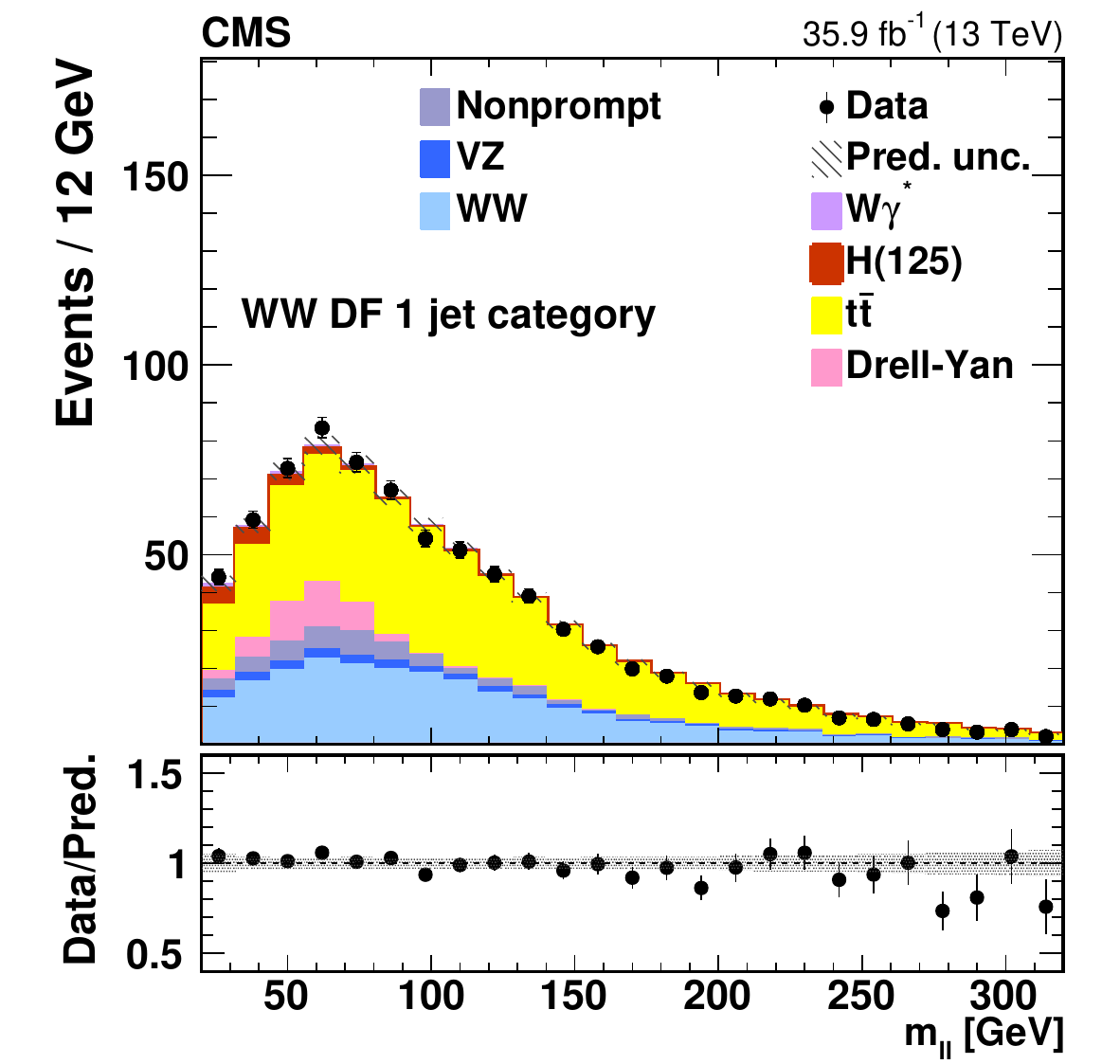}
\caption[.]{\label{fig:trad1jet}
Kinematic distributions for events with exactly one jet and DF leptons
in the sequential cut analysis.
The quantities, error bars, and hatched areas are the same as in Fig.~\ref{fig:trad0jet}.}
\end{figure}

Table~\ref{tab:event_sel_both} summarizes the event selection criteria and
Table~\ref{tab:yields} lists the sample composition after the fits described
in Section~\ref{sec:signal} have been executed.
Example kinematic distributions are shown in Fig.~\ref{fig:trad0jet} for
events with no jets and in Fig.~\ref{fig:trad1jet} for events with
exactly one jet. The simulations reproduce the observed distributions well.

\begin{table*}[ht]
  \centering
  \topcaption{Summary of the event selection criteria for the sequential cut and
the random forest analyses.
DYMVA refers to an event classifier used in the sequential cut analysis to suppress
Drell--Yan background events.  RF refers to random forest classifiers.
Kinematic quantities are measured in\unit{GeV}.  The symbol (\NA) means no requirement applied.}
  \begin{scotch}{ l cc cc }
 Quantity & \multicolumn{2}{c}{Sequential Cut} & \multicolumn{2}{c}{Random Forest} \\
 & DF & SF & DF & SF \\
  \hline
Number of leptons             & \multicolumn{2}{c}{Strictly 2}  & \multicolumn{2}{c}{Strictly 2} \\
Lepton charges                & \multicolumn{2}{c}{Opposite}    & \multicolumn{2}{c}{Opposite}  \\
$\ptmax$                      & \multicolumn{2}{c}{$>$25}       & \multicolumn{2}{c}{$>$25} \\
$\ptmin$                      & \multicolumn{2}{c}{$>$20}       & \multicolumn{2}{c}{$>$20} \\
$\mll$                        & $>$20        & $>$40        & $>$30    & $>$30 \\
Additional leptons            & \multicolumn{2}{c}{0}       & \multicolumn{2}{c}{0}    \\
$\abs{\mll-m_{\PZ}}$           & \NA          & $>$15        & \NA      & $>$15 \\
$\qT$                         & $>$30        & $>$30        & \NA      & \NA \\
\ptmiss                              & $>$20  		& $>$55   & \NA & \NA \\
$\Etmissproj$, $\Mettrackproj$         & $>$20                  & $>$20   & \NA & \NA \\
Number of jets                         & \multicolumn{2}{c}{$\le$1} & \NA & \NA \\
Number of \PQb-tagged jets             & \multicolumn{2}{c}{0} & \multicolumn{2}{c}{0} \\
DYMVA   score                          & \NA                       & $>$0.9  & \NA & \NA \\
Drell--Yan RF score $\SDY$             & \NA & \NA & \multicolumn{2}{c}{$>$0.96} \\
\ttbar RF score $\STT$                 & \NA & \NA & \multicolumn{2}{c}{$>$0.6} \\
  \end{scotch}
\label{tab:event_sel_both}
\end{table*}

\begin{table*}
  \centering      
  \topcaption{Sample composition for the sequential cut and random forest selections after
  the fits described in Section~\ref{sec:signal} have been executed; the uncertainties shown
  are based on the total uncertainty obtained from the fit.
  The purity is the fraction of selected events that are \WWpm signal events.
  ``Observed'' refers to the number of events observed in the data. }
  \newcolumntype{x}{D{,}{\,\pm\,}{-1}}
\cmsTable{
\begin{scotch}{ l   x x x x    x x }
Process & \multicolumn{4}{c}{Sequential Cut} & \multicolumn{2}{c}{Random Forest}  \\
 & \multicolumn{2}{c}{DF} & \multicolumn{2}{c}{SF} & \multicolumn{1}{c}{DF} & \multicolumn{1}{c}{SF} \\
 & \multicolumn{1}{c}{0-jet} & \multicolumn{1}{c}{1-jet} & \multicolumn{1}{c}{0-jet} & \multicolumn{1}{c}{1-jet}  & \multicolumn{2}{c}{all jet multiplicities}  \\
\hline
 \multicolumn{7}{c}{ } \\[-1.5ex]
Top quark            &  2110,110   & 5000,120  & 1202,66    & 2211,69         & 3450,340      & 830,82   \\ 
Drell--Yan           &   129,10    &  498,38   & 1230,260   &  285,86         & 1360,130      & 692,72   \\ 
$\PV\PZ$             &   227,13    &  270,12   &  192,12    &  110,7          &  279,29       & 139,10   \\ 
\PV\PV\PV            &    11,1     &   29,2    &    4,1     &    6,1          &   13,4        &   3,2    \\
$\PH\to\WWpm$        &   269,41    &  150,25   &   50,2     &   27,1          &  241,26       &  90,10   \\ 
\PW$\gamma^{(*)}$     &   147,17    &  136,13   &  123,5     &   58,6          &  305,88       &  20,6   \\ 
Nonprompt leptons    &   980,230   &  550,120  &  153,39    &  127,32         &  940,300      & 183,59   \\
Total background     &  3870,260   & 6640,180  & 2950,270   & 2820,120        &  \multicolumn{2}{c}{ }\\
      	& \multicolumn{2}{x}{10\,510,310}   & \multicolumn{2}{x}{5780,300}    & 6600,480      & 1960,120 \\[1ex]
$\qqbar \to \WWpm$   &  6430,250   & 2530,140  & 2500,180   & 1018,71         & 12\,070,770   & 2820,180 \\ 
$\Pg\Pg \to \WWpm$   &   521,66    &  291,38   &  228,32    &  117,15         &     693,44    &  276,17   \\ 
Total $\WWpm$        &  6950,260   & 2820,150  & 2730,190   & 1136,72         & \multicolumn{2}{c}{ } \\
        & \multicolumn{2}{x}{9780,300} & \multicolumn{2}{x}{3860,200}         &  12\,770,820  &  3100,200 \\[1ex]
Total yield          & 10\,820,360 & 9460,240  & 5680,330   & 3960,360        & \multicolumn{2}{c}{ } \\
        & \multicolumn{2}{x}{20\,280,430} & \multicolumn{2}{x}{9640,490}      & 19\,360,950   &  5060,240 \\[1ex]
Purity               & \multicolumn{1}{c}{0.64} & \multicolumn{1}{c}{0.30} & \multicolumn{1}{c}{0.48} & \multicolumn{1}{c}{0.29} & \multicolumn{2}{c}{ } \\
                     & \multicolumn{2}{c}{0.48} & \multicolumn{2}{c}{0.40} & \multicolumn{1}{c}{0.66} & \multicolumn{1}{c}{0.61} \\[1.5ex]
Observed             & \multicolumn{1}{c}{10\,866} & \multicolumn{1}{c}{9404} & \multicolumn{1}{c}{5690} & \multicolumn{1}{c}{3914} & \multicolumn{1}{c}{19\,418} & \multicolumn{1}{c}{5210}\\[1ex]
\end{scotch}
}
\label{tab:yields}
\end{table*}

\subsection{Random forest selection}
\label{sec:selectionRF}

A random forest (RF) classifier is an aggregate of binary decision trees that have been trained
independently and in parallel~\cite{Breiman}.  Each individual tree uses a random subset of features
which mitigates against overfitting, a problem that challenges other classifiers based on decision trees.
The random forest classifier is effective if there are many trees, and the aggregation of many trees
averages out potential overfitting by individual trees.
A random forest classifier is expected to improve monotonically without overfitting~\cite{Ho1998}
in contrast to other methods.  Building a random
forest classifier requires less tuning of hyperparameters compared, for example, with
boosted decision trees, and its performance is as good~\cite{Caruana2005}.

The random forest analysis begins with a preselection that is close to
the first set of requirements in the sequential cut analysis.  The selection
of electrons and muons is identical.  To avoid low-mass resonances and leptons
from decays of hadrons, $\mll>30\GeV$ is required for both DF
and SF events.  To suppress the large background contribution from
$\PZ$~boson decays, events with SF leptons and with $\mll$ within 15\GeV of
the $\PZ$~boson mass are rejected.
Events with a third, loosely identified lepton with $\pt>10\GeV$ are
rejected to reduce backgrounds from $\PV\PZ$ production.
Finally, events with one or more \PQb-tagged jets ($\pt^{\PQb}>20\GeV$ and medium working point) are rejected,
since the background from \ttbar production is characterized by the presence of \PQb jets whereas the
signal is not.   These requirements are known as the preselection requirements.

\begin{figure}
\centering
\includegraphics[width=\cmsFigWidthTwo]{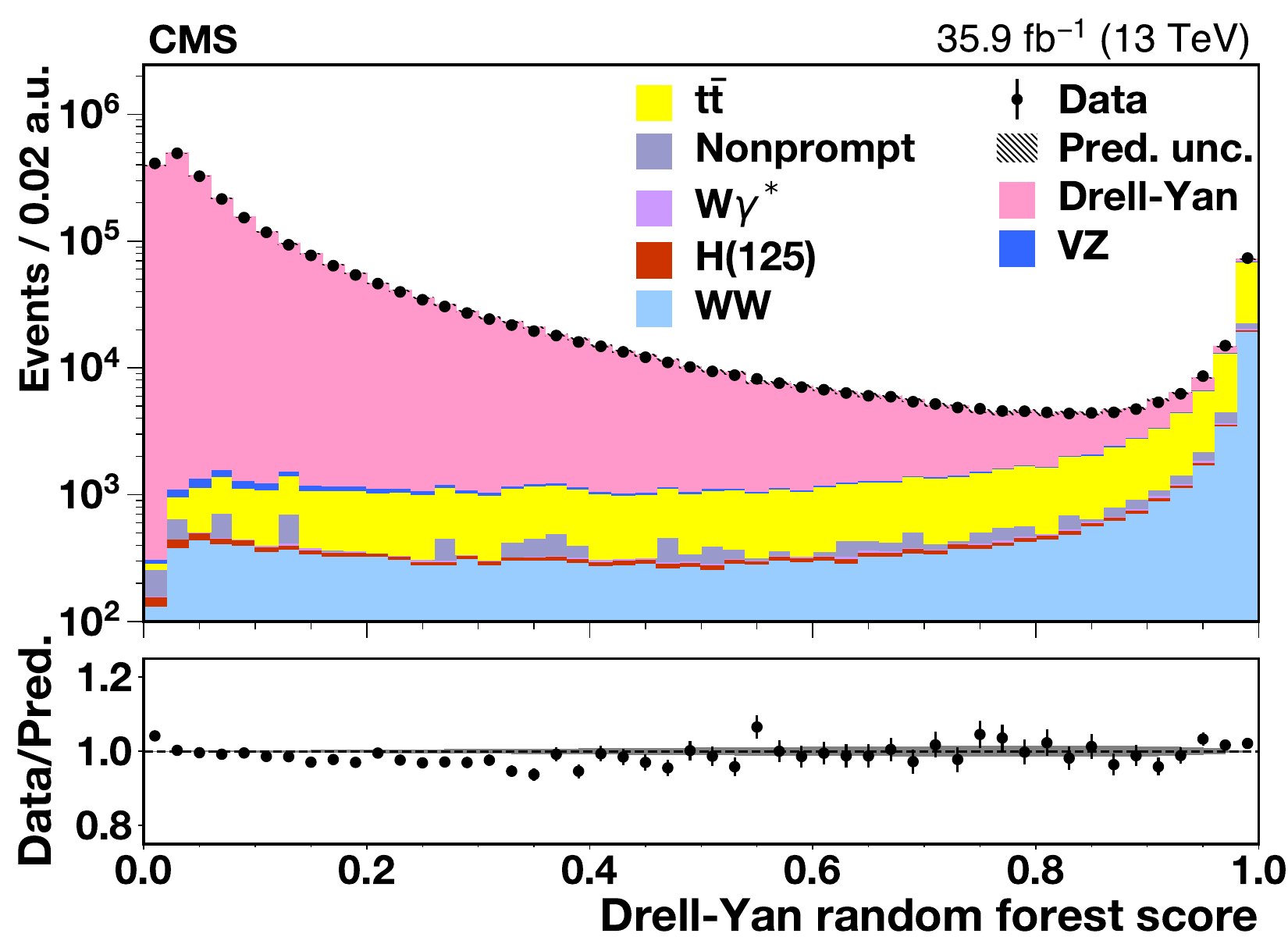}
\includegraphics[width=\cmsFigWidthTwo]{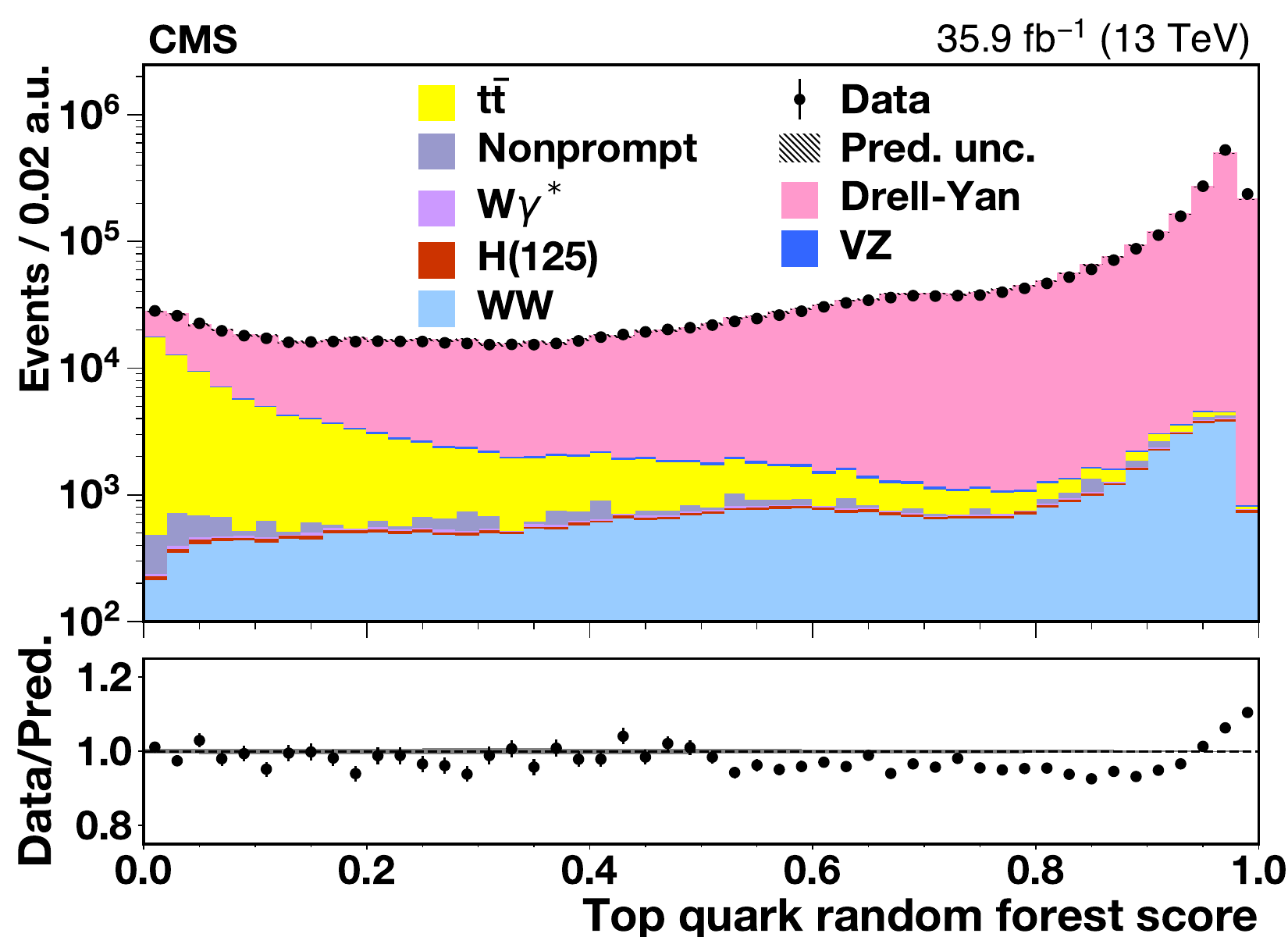}\\
\includegraphics[width=\cmsFigWidthTwo]{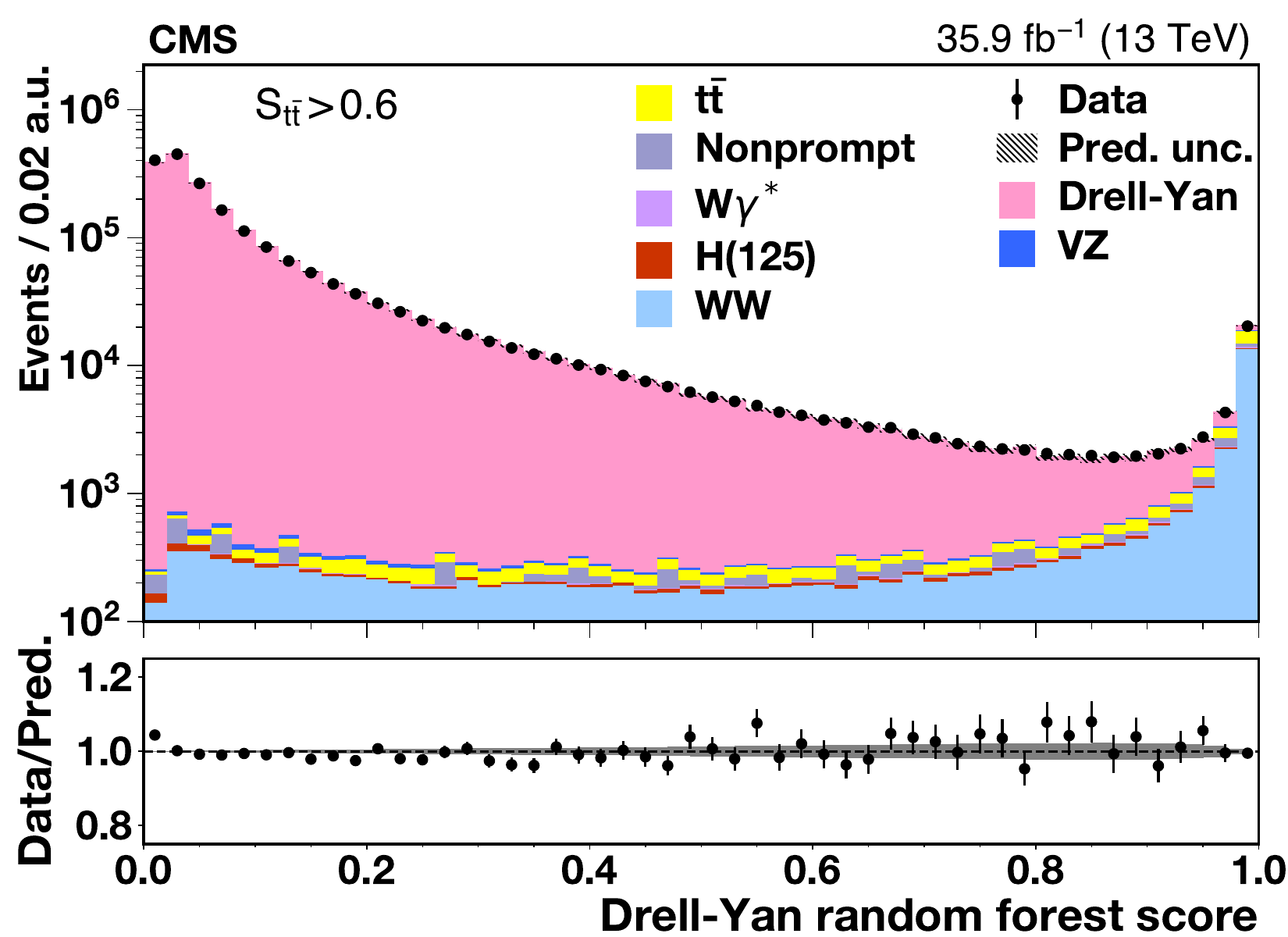}
\includegraphics[width=\cmsFigWidthTwo]{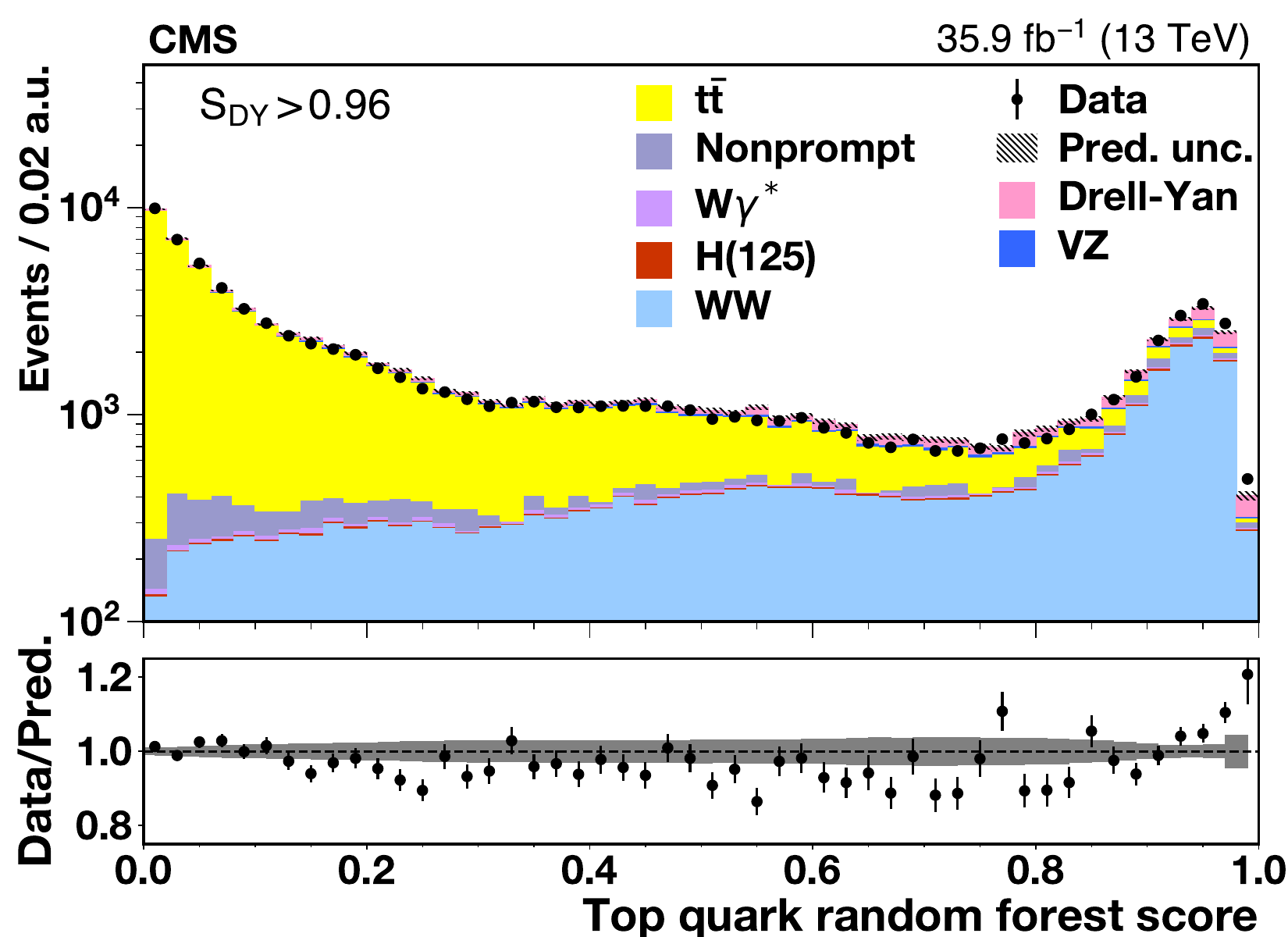}
\caption[.]{\label{fig:scores}
Top left: score $\SDY$ distribution for the Drell--Yan discriminating random forest discriminant.
The Drell--Yan  distribution peaks toward zero and the \WWpm distribution peaks toward one.
Top right: score $\STT$ distribution for the top quark random forest discriminant.
The \ttbar distribution peaks toward zero and the \WWpm peaks toward one.
Bottom left: the $\SDY$ distribution after suppressing top quark events with $\STT>\STTmin=0.6$.
Bottom right: the $\STT$ distribution after suppressing Drell--Yan events with $\SDY>\SDYmin=0.96$.
The error bars on the points represent the statistical uncertainties for the data,
and the hatched areas represent the combined systematic and statistical uncertainties
of the predicted yield in each bin.}
\end{figure}

After the preselection, the largest background contamination comes from Drell--Yan production
of lepton pairs and \ttbar production with both top quarks producing prompt leptons.
To reduce these backgrounds, two independent random forest classifiers are 
constructed: an anti-Drell--Yan classifier optimized to distinguish Drell--Yan and
\WWpm signal events, and an anti-$\ttbar$ classifier optimized to distinguish \ttbar
and \WWpm events.  The classifiers produce scores, $\SDY$ and $\STT$, arranged so
that signal appears mainly at $\SDY\approx1$ and $\STT\approx1$ while backgrounds
appear mainly at $\SDY\approx0$ and $\STT\approx0$.  Figure~\ref{fig:scores} shows
the distributions of the scores for the two random forest classifiers.
The signal region is defined by the requirements $\SDY > \SDYmin$ and $\STT > \STTmin$.
For the cross section measurement, the specific values $\SDYmin=0.96$ and $\STTmin=0.6$
are set by simultaneously minimizing the uncertainty in the cross section and maximizing
the purity of the selected sample.  For measuring the jet multiplicity, a lower
value of $\STTmin=0.2$ is used, which increases the efficiency for \WWpm events
with jets.  A Drell--Yan control region is defined by $\SDY<0.6$ and $\STT>0.6$ 
and a \ttbar control region is defined by $\SDY>0.6$ and $\STT<0.6$.
The event selection used in this measurement is summarized in 
Table~\ref{tab:event_sel_both}.

The architecture of the two random forest classifiers is determined
through an optimization of hyperparameters explored in a grid-like fashion.
The optimal architecture for this problem has 50 trees with a maximum tree
depth of 20; the minimum number of samples per split is 50 and the minimum
number of samples for a leaf is one.  The maximum number of features seen
by any single tree is the square-root of the total number of features (ten for the
DY random forest and eight for the \ttbar random forest).

The random forest classifier takes as input some of the kinematic quantities listed
in Table~\ref{tab:event_sel_both} and several other event features as listed in Table~\ref{tab:feature_list}.
These include the invariant mass of the two leptons and the missing
momentum vector $m_{\Pell\Pell\ptmiss}$, the azimuthal angle between the lepton pair
and the missing momentum vector $\Delta\phi_{\Pell\Pell\ptmiss}$, the smallest azimuthal
angle between either lepton and any reconstructed jet $\Delta\phi_{\Pell\text{J}}$,
and the smallest azimuthal angle between the missing momentum vector and any jet
$\Delta\phi_{\ptmiss\text{J}}$.   The random forest classifier also makes use
of the scalar sum of jet transverse momenta $\HT$, and of the vector
sum of the jet transverse momenta, referred to as the recoil in the event.

\begin{table}
\centering
\topcaption{	\label{tab:feature_list}
Features used for the random forest classifiers.  The first classifier
distinguishes Drell--Yan and \WWpm signal events, and the second one distinguishes
top quark events and signal events.}
\begin{scotch}{ l  c c }
 Feature & \multicolumn{2}{c}{Classifier} \\
  & Drell--Yan 	& Top quark\\ 
\hline
Lepton flavor	      &\checkmark	&	\\
Number of jets	      & 		      & \checkmark\\
$\ptmin$              &\checkmark	& \\
\ptmiss               &\checkmark	&\checkmark \\
$\Etmissproj$         &\checkmark	& \\
$\qT$		      &\checkmark	&\checkmark \\
$\mll$		      &\checkmark	& \\
$m_{\Pell\Pell\ptmiss}$  &\checkmark	& \\
$\Delta\phi_{\Pell\Pell\ptmiss}$	 &\checkmark	& \checkmark \\
$\Delta\phi_{\Pell\text{J}}$    & 		      &\checkmark \\
$\Delta\phi_{\ptmiss\text{J}}$    & 		      &\checkmark \\
$\Delta\phi_{\Pell\Pell}$	    &\checkmark	& \\
$\HT$		            & 		& \checkmark \\
Recoil		            &\checkmark	& \checkmark \\
\end{scotch}
\end{table}

The sample composition for the signal region is summarized in Table~\ref{tab:yields}.
The signal efficiency and purity are higher than in the sequential cut analysis.

\section{Background estimation}\label{sec:backgrounds}
A combination of methods based on data control samples and simulations are used to 
estimate background contributions. The methods used in the sequential cut analysis and
the random forest analysis are similar.   The differences are described below.

The largest background contribution comes from \ttbar and single top production which
together are referred to as top quark production.  This contribution arises when
\PQb~jets are not tagged either because they fall outside the kinematic  region where
tagging is possible or because they receive low scores from the \CSVvv \PQb tagging
algorithm.   The sequential cut analysis defines a control region by requiring at least one \PQb-tagged jet.
The normalization of the top quark background in the signal region is set according to the
number of events in this control region.  Similarly, the random forest analysis
defines a top quark control region on the basis of scores: $\SDY>0.6$ and
$\STT<0.6$.  Many kinematic distributions are examined and all show good agreement
between simulation and data in this control region.  This control region is used to set the normalization
of the top quark background in the signal region.

The next largest background contribution comes from the Drell--Yan process, which is
larger in the SF channel than in the DF channel.
The nature of these contributions is somewhat different.  The SF contribution
arises mainly from the portion of Drell--Yan production that falls below or above
the \PZ resonance peak.  The sequential analysis calibrates this contribution
using the observed number of events in the \PZ peak and the ratio of numbers of
events inside and outside the peak, as estimated from simulations.
The DF contribution arises from $\PZ\to\tau^+\tau^-$ production with
both $\tau$ leptons decaying leptonically.  The sequential cut analysis verifies the
$\PZ\to\tau^+\tau^-$ background using a control region defined by
$\memu<80\GeV$ and inverted $\qT$ requirements.
The random forest analysis defines a
Drell--Yan control region by $\SDY<0.6$ and $\STT>0.6$, which includes
both SF and DF events. Simulations of kinematic distributions for events in this region
match the data well, and the yield of events in this region is used to normalize the Drell--Yan
background contribution in the signal region.

The next most important background contribution comes mainly from \PW boson events
in which a nonprompt lepton from a jet is selected in addition to a lepton from the
\PW boson decay.  Monte Carlo simulation cannot be used for an accurate estimate
of this contribution, but it can be used to devise and evaluate an estimate based on control samples.
In the sequential cut analysis, a ``pass-fail'' control sample is defined by one lepton that
passes the lepton selection criteria and another that fails the criteria but passes 
looser criteria.  The misidentification rate $f$ for a jet that satisfies loose lepton requirements
to also pass the standard lepton requirements is determined using an event sample
dominated by multijet events with nonprompt leptons.   This misidentification
rate is parameterized as a function of lepton \pt and $\eta$ and used to compute
weights $f/(1-f)$ in the pass-fail sample that are used to determine the contribution
of nonprompt leptons in the signal region~\cite{Sirunyan:2018egh,HiggsWW2016}.
The random forest analysis uses a different method based on a control region
in which the two leptons have the same charge.  This control region is dominated
by \PW{+}jets events with contributions from diboson and other events.  The transfer factor relating
the number of same-sign events in the control region to the number of opposite-sign events
in the signal region is based on two methods relying on data control samples and which are validated using simulations.
One method uses events with DF leptons and low \ptmiss and the other uses
events with an inverted isolation requirement.  Both methods yield values for the
transfer factor that are consistent at the 16\% level.

Background contamination from $\PW\gamma^{\ast}$ events with low-mass $\gamma^{\ast}\to\Pell^+\Pell^-$  
can satisfy the signal event selection when the transverse momenta of the two leptons are 
very different~\cite{HiggsWW2016}. The predicted contribution in the signal region
is normalized to the number of events in a control region with three muons 
satisfying $\pt>10$, 5, and 3\GeV and for $m_{\gamma^{\ast}}<4\GeV$. 
In this control region, the requirement $\ptmiss<25\GeV$ is imposed
in order to suppress non-$\PW\gamma^\ast$ events.

The remaining minor sources of background, including diboson and triboson final states
and Higgs-mediated \WWpm production, are evaluated using simulations normalized to
the most precise theoretical cross sections available.

\section{Signal extraction}\label{sec:signal}
The cross sections are obtained by simultaneously fitting the predicted yields to the observed yields
in the signal and control regions.
In this fit, a signal strength parameter modifies the predicted signal yield
defined by the central value of the theoretical cross section, $\sigmaNNLO = \sigmaNNLOval$.
The fitted value of the signal strength is expected to be close to unity if the SM is valid,
and the measured cross section is the product of the signal strength and
the theoretical cross section.  Information from control regions is incorporated
in the analysis through additional parameters that are free in the fit;
the predicted background in the signal region is thereby tied to the yields
in the control regions.   In the sequential cut analysis, there is one control
region enriched in \ttbar events; the yields in the signal and this one control
region are fit simultaneously.  Since the selected event sample is separated
according to SF and DF, 0- and 1-jet selections, there are
eight fitted yields.   In the random forest analysis there are
three control regions, one for Drell--Yan background, a second for \ttbar
background, and a third for events with nonprompt leptons (\eg, \PW{+}jets). 
Since SF and DF final states are analyzed together, and the
selection does not explicitly distinguish the number of jets, there
are four fitted yields in the random forest analysis.
In both analyses, the yields in the control regions effectively constrain
the predicted backgrounds in the signal regions.

Additional nuisance parameters are introduced in the fit that encapsulate important sources
of systematic uncertainty, including the electron and muon efficiencies,
\PQb tagging efficiencies, the jet energy scale, 
and the predicted individual contributions to the background.
The total signal strength uncertainty, including all systematic uncertainties,
is determined by the fit with all parameters free;
the statistical uncertainty is determined by fixing all parameters except
the signal strength to their optimal values.

\section{Systematic uncertainties}\label{sec:systematics}
Experimental and theoretical sources of systematic uncertainty
are described in this section.  A summary of all systematic
uncertainties for the cross section measurement is given in 
Table~\ref{tab:wwSystematics}.  These sources of uncertainty
impact the measurements of the cross section through the normalization
of the signal.  Many of them also impact kinematic distributions
that ultimately can alter the shapes of distributions studied in
this analysis.  Both normalization and shape uncertainties are evaluated.

\subsection{Experimental sources of uncertainty}

There are several sources of experimental systematic uncertainties, including the lepton 
efficiencies, the \PQb-tagging efficiency for \PQb~quark jets and the mistag rate
for light-flavor quark and gluon jets,
the lepton momentum and energy scales, the jet energy scale and resolution,
the modeling of \ptmiss and of pileup in the simulation,
the background contributions, and the integrated luminosity.

The sequential cut and the random forest analyses both use control regions to estimate the background
contributions in the signal region.  The uncertainties in the estimates are
determined mainly by the statistical power of the control regions, though the
uncertainty of the theoretical cross sections and the shape of the \PZ resonant
peak also play a role.  
Sources of systematic uncertainty of the estimated Drell--Yan background
include the \PZ resonance line shape and the performance of the DYMVA classifier for
different \ptmiss thresholds.  These uncertainties are propagated directly to the
predicted SF and DF background estimates.
The contribution from nonprompt leptons is entirely
determined by the methods based on data control regions, described in Section~\ref{sec:backgrounds};
typically these contributions are uncertain at approximately the 30\% level.
The contribution from the $\PW\gamma^\ast$ final state is checked using a sample of events with
three well-identified leptons including a low-mass, opposite-sign pair of muons.
The comparison of the MC prediction with the data has an uncertainty of about 20\%.
The other backgrounds are estimated using simulations and their uncertainties
depend on the uncertainties of the theoretical cross sections, which are typically
below 10\%.   Statistical uncertainties from the limited number of MC
events are taken into account, and have a very small impact on the result.

Small differences in the lepton trigger, reconstruction, and identification efficiencies
for data and simulation are corrected by applying scale factors to adjust
the efficiencies in the simulation.  These scale factors are obtained using
events in the \cPZ resonance peak region~\cite{Khachatryan:2015hwa, Chatrchyan:2013sba}
recorded with unbiased triggers. They vary with lepton \pt 
and $\eta$ and are within 3\% of unity.  The uncertainties of these scale factors
are mostly at the 1--2\% level.

Differences in the probabilities for \PQb~jets and light-flavor quark and gluon jets
to be tagged by the \CSVvv algorithm are corrected by applying scale factors
to the simulation.    These scale factors are measured using 
\ttbar events with two leptons~\cite{Chatrchyan:2012jua}.  These scale factors
are uncertain at the percent level and have relatively little impact on
the result because the signal includes mainly light-flavor quark and gluon jets, which have
a low probability to be tagged, and the top quark background is assessed
using appropriate control regions.

The jet energy scale is set using a variety of in situ calibration techniques~\cite{JES}.
The remaining uncertainty is assessed as a function of jet \pt and $\eta$.
The jet energy resolution in simulated events is slightly different than that
measured in data.  The differences between simulation and data lead to
uncertainties in the efficiency of the event selection because the number
of selected jets, their transverse momenta, and also \ptmiss play a role in the event
selection.

The lepton energy scales are set using the position of the \PZ resonance peak;
the uncertainties are very small and have a negligible impact on the measurements
reported here.

The modeling of pileup depends on the total inelastic $\Pp\Pp$ cross section~\cite{Sirunyan:2018nqx}.
The pileup uncertainty is evaluated by varying this cross section up and down by 5\%.

The statistical uncertainties from the limited number of events in the
various control regions lead to a systematic uncertainty from the background
predictions.   It is listed as part of the experimental systematic
uncertainty in Table~\ref{tab:wwSystematics}.

The uncertainty in the integrated luminosity measurement is 2.5\%~\cite{REFLUMI}. 
It contributes directly to the cross section and also to the uncertainty in
the minor backgrounds predicted from simulation.

\subsection{Theoretical sources of uncertainty}
\label{sec:unctheory}

The efficiency of the event selection is sensitive to the number of hadronic jets in the event.
The sequential cut analysis explicitly singles out events with zero or one jet, and
the random forest classifiers utilize quantities, such as $\HT$, that tend to
correlate with the number of jets.   As a consequence, the efficiency of the
event selection is sensitive to higher-order QCD corrections that are adequately
described by neither the matrix-element calculation of \POWHEG nor by the parton shower simulation.
The uncertainty reflecting these missing higher orders is evaluated by varying the
QCD factorization and renormalization scales independently up and down by a factor of two but excluding
cases in which one is increased and the other decreased simultaneously.
A change in measured cross sections is evaluated by applying appropriate weights to the simulated events.

Some of the higher-order QCD contributions to \WWpm production have been calculated
using the $\pt$-resummation~\cite{wwptnnll,wwpt} and the jet-veto resummation~\cite{wwjv} techniques. 
The results from these two approaches are compatible~\cite{wwptvj}.
The transverse momentum \WWpt of the \WWpm pair is used as a proxy for these
higher-order corrections; the \WWpt spectrum from \POWHEG is reweighted to match the
analytical prediction obtained using the $\pt$-resummation at next-to-next-to-leading logarithmic
accuracy~\cite{wwptnnll}.
Uncertainties in the theoretical calculation of the \WWpt spectrum lead to
uncertainties in the event selection efficiency that are assessed
for the $\Pq\Paq\to\WWpm$ process by independently varying the
resummation, the factorization, and the renormalization scales in the analytical
calculation~\cite{wwpt}.  The uncertainty in the $\Pg\Pg\to\WWpm$
component is determined by the variation of the renormalization and factorization scales
in the theoretical calculation of this process~\cite{Caola:2015rqy}.

Additional sources of theoretical uncertainties come from the PDFs and the
assumed value of~$\alpS$.  The PDF uncertainties are estimated, following the PDF4LHC 
recommendations~\cite{Butterworth:2015oua}, from the variance of the values
obtained using the set of MC replicas of the NNPDF3.0 PDF set.
The variation of both the signal and the backgrounds with each PDF set
and the value of $\alpS$ is taken into account.

The uncertainty from the modeling of the underlying event 
is estimated by comparing the signal efficiency obtained with the $\Pq\Paq\to\WWpm$ 
sample described in Section~\ref{sec:samples} to alternative samples that use
different generator configurations.

The branching fraction for leptonic decays of \PW bosons is taken to be
$\BFWlept=0.1086\pm0.0009$~\cite{PDG2018}, and lepton universality
is assumed to hold.   The uncertainty coming from this branching fraction
is not included in the total uncertainty; it would amount to 1.8\% of the cross section value.

\begin{table}[ht]
\centering
\topcaption{\label{tab:wwSystematics}
Relative systematic uncertainties in the total cross section measurement
(0- and 1-jet,  DF and SF) based on the sequential cut analysis.}
\begin{scotch}{ l c }
Uncertainty source &  (\%) \\
  \hline
 & \\[-2ex]
Statistical                                     & 1.2    \\[\cmsTabSkip]
\ttbar normalization                              & 2.0    \\
Drell--Yan normalization                          & 1.4    \\
$\PW\gamma^\ast$ normalization                     & 0.4    \\         
Nonprompt leptons normalization                   & 1.9    \\
Lepton efficiencies                               & 2.1    \\
\PQb tagging (\PQb/\PQc)                          & 0.4    \\
Mistag rate (\cPq/\Pg)                            & 1.0    \\
Jet energy scale and resolution                   & 2.3    \\
Pileup                                            & 0.4    \\
Simulation and data control regions sample size   & 1.0    \\
Total experimental systematic                     & 4.6    \\[\cmsTabSkip]
QCD factorization and renormalization scales      & 0.4    \\
Higher-order QCD corrections and $\WWpt$ distribution & 1.4    \\
PDF and $\alpS$                                   & 0.4    \\
Underlying event modeling                         & 0.5    \\
Total theoretical systematic                      & 1.6    \\[\cmsTabSkip]
Integrated luminosity                              & 2.7    \\[\cmsTabSkip]
Total                                             & 5.7    \\[\cmsTabSkip] 
\end{scotch}
\end{table}

\section{The \texorpdfstring{$\WWpm$}{W+W-} cross section measurements}\label{sec:WWxsec}
Two measurements of the total production cross section are reported in this section:
the primary one coming from the sequential cut analysis and a secondary measurement
coming from the random forest analysis.   In addition, measurements of the fiducial
cross section are reported, based on the sequential cut analysis, including
the change of the zero-jet cross section with variations of the jet \pt threshold.

\subsection{Total production cross section}
\label{sec:TotalCrossSection}

Both the sequential cut and random forest analyses provide
precise measurements of the total production cross section.
Since the techniques for selecting signal events are rather different, both
values are reported here.  The measurement obtained with the sequential
cut analysis is the primary measurement of the total production cross section because
it is relatively insensitive to the uncertainties in the corrections applied
to the \WWpt spectrum.
The overlap of the two sets of selected events is approximately~50\%.
A combination of the two measurements is not carried out because the reduction
in the uncertainty would be minor.

The sequential cut (SC) analysis makes a double-dichotomy of the data: selected events are separated
if the leptons are DF or SF (DF is purer because of a
smaller Drell-Yan contamination), and these are further subdivided depending
on whether there is zero or one jet (0-jet is purer because of a smaller top quark contamination).
The comparison of the four signal strengths provides an important test of the
consistency of the measurement; the cross section value is based on the
simultaneous fit of DF \& SF and 0-jet \& 1-jet channels.
The result is
$\sigmatotTA = 117.6 \pm 1.4\stat \pm 5.5\syst \pm 1.9\thy \pm 3.2\lum\pb = 117.6 \pm 6.8\pb$,
which is consistent with the theoretical prediction  $\sigmaNNLO = \sigmaNNLOval$.
A summary of the measured signal strengths and the corresponding cross sections is given in Table~\ref{tab:TAsignalstrengths}.

\begin{table}[tph]
\centering
\topcaption[.]{\label{tab:TAsignalstrengths}
Summary of the signal strength and total production cross section obtained in the sequential cut analysis.
The uncertainty listed is the total uncertainty obtained from the fit to the yields.}
\begin{scotch}{ l l  c  c  }
\multicolumn{2}{ c }{Category} & Signal strength & Cross section [pb] \\
\hline
0-jet & DF                & $1.054 \pm 0.083$ &  $125.2 \pm \phantom{0}9.9$ \\
0-jet & SF                & $1.01\phantom{1} \pm 0.16\phantom{0}$ &  $120\phantom{.1} \pm 19\phantom{.0}$ \\
1-jet & DF                & $0.93\phantom{0} \pm 0.12\phantom{4}$ &  $110\phantom{.5} \pm 15\phantom{.7}$ \\
1-jet & SF                & $0.76\phantom{0} \pm 0.20\phantom{0}$ &  $\phantom{0}89\phantom{.9} \pm 24\phantom{.8}$ \\
0-jet \& 1-jet & DF       & $1.027 \pm 0.071$ &  $122.0 \pm \phantom{0}8.4$ \\
0-jet \& 1-jet & SF       & $0.89\phantom{2} \pm 0.16\phantom{7}$ &  $106\phantom{.0} \pm 19\phantom{.7}$ \\
0-jet \& 1-jet & DF \& SF & $0.990 \pm 0.057$ &  $117.6 \pm \phantom{0}6.8$ \\
\end{scotch}
\end{table}

\begin{figure}
\centering
\includegraphics[width=0.45\textwidth]{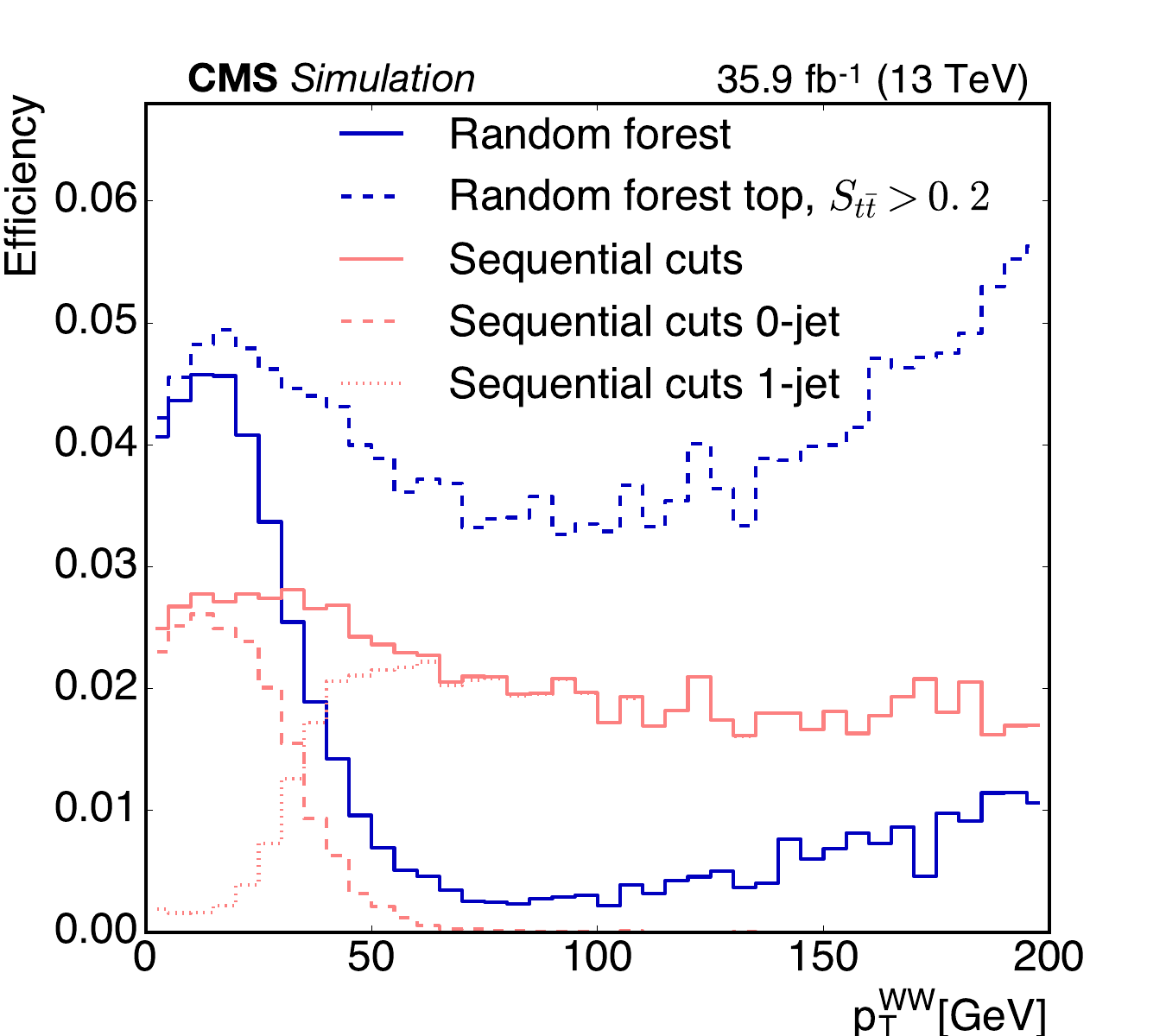}
\caption[.]{\label{fig:WWptcompare}
Comparison of efficiencies for the sequential cut and random forest analyses
as a function of \WWpt.   The sequential cut analysis includes 0- and 1-jet
events from both DF and SF lepton combinations, for which the contributions from
0- and 1-jet are shown separately.   The efficiency curve for $\STTmin = 0.2$
is also shown; this value is used in measuring the jet multiplicity distribution.}
\end{figure}

The random forest analysis isolates a purer signal than the sequential cut analysis
(see Table~\ref{tab:yields}); however, its sensitivity is concentrated at relatively low \WWpt
as shown in Fig.~\ref{fig:WWptcompare}.  This region corresponds mainly to
events with zero jets; the random forest classifier uses observables such as $\HT$
that correlate with jet multiplicity and reduce top quark background
contamination by favoring events with a low jet multiplicity.   As a consequence,
the random forest result is more sensitive to uncertainties in the theoretical corrections to
the \WWpt spectrum than the sequential cut analysis. The signal strength measured by the random
forest analysis is $1.106\pm0.073$ which corresponds to a measured total production cross section of
$\sigmatotRF = 131.4 \pm 1.3\stat \pm 6.0\syst \pm 5.1\thy \pm 3.5\lum\pb = 131.4 \pm 8.7\pb$.
The difference with respect to the sequential cut analysis reflects the sensitivity of the
random forest analysis to low~\WWpt.

\subsection{Fiducial cross sections}
\label{sec:fiducialX}
The sequential cut analysis is used to obtain fiducial cross sections.
The definition of the fiducial region is similar to the requirements described
in Section~\ref{sec:selectionTA} above.  The generated event record must contain
two prompt leptons (electrons or muons) with $\pt>20\GeV$ and $\abs{\eta}<2.5$.
Decay products of $\tau$ leptons are not considered part of the signal in
this definition of the fiducial region.  Other kinematic requirements are applied:
$\mll>20\GeV$, $\qT>30\GeV$, and $\ptmiss>20\GeV$ (where \ptmiss is calculated using
the momenta of the neutrinos emitted in the \PW\ boson decays). When categorizing
events with zero or more jets, a jet is defined using stable particles but not neutrinos.
For the baseline measurements, the jets must have $\pt>30\GeV$ and $\abs{\eta} < 4.5$
and be separated from each of the two leptons by $\Delta R > 0.4$.

The fiducial cross section is obtained by means of a simultaneous fit to the
DF and SF, 0- and 1-jet final states. The measured value is
$\sigmafidTA = 1.529 \pm 0.020\stat \pm 0.069\syst \pm 0.028\thy \pm 0.041\lum\pb
= 1.529 \pm 0.087\pb$, which agrees well with 
the theoretical value $\sigma_{\text{NNLO}}^{\text{fid}} = 1.531 \pm 0.043\pb$.
These values are corrected to the fiducial region with all jet multiplicities.

The fiducial cross sections for the production of \WWpm boson pairs with zero or one jet are
of interest because some of the earlier measurements were based on the 0-jet subset only,
\ie, a jet veto was applied~\cite{ATLAS:2012mec,Aad:2016wpd,Chatrchyan:2013yaa,Aaltonen:2009aa}.
The sequential cut analysis provides the following values based on the combination
of the DF and SF categories:
$\sigmafidTA({\text{0-jet}}) = 1.61\pm0.10\pb$ and
$\sigmafidTA({\text{1-jet}}) = 1.35\pm0.11\pb$ for a jet \pt threshold of 30\GeV.
These fiducial cross section values pertain to the definition given above, in particular,
they pertain to all jet multiplicities.

\begin{figure}
\centering
\includegraphics[width=0.45\textwidth]{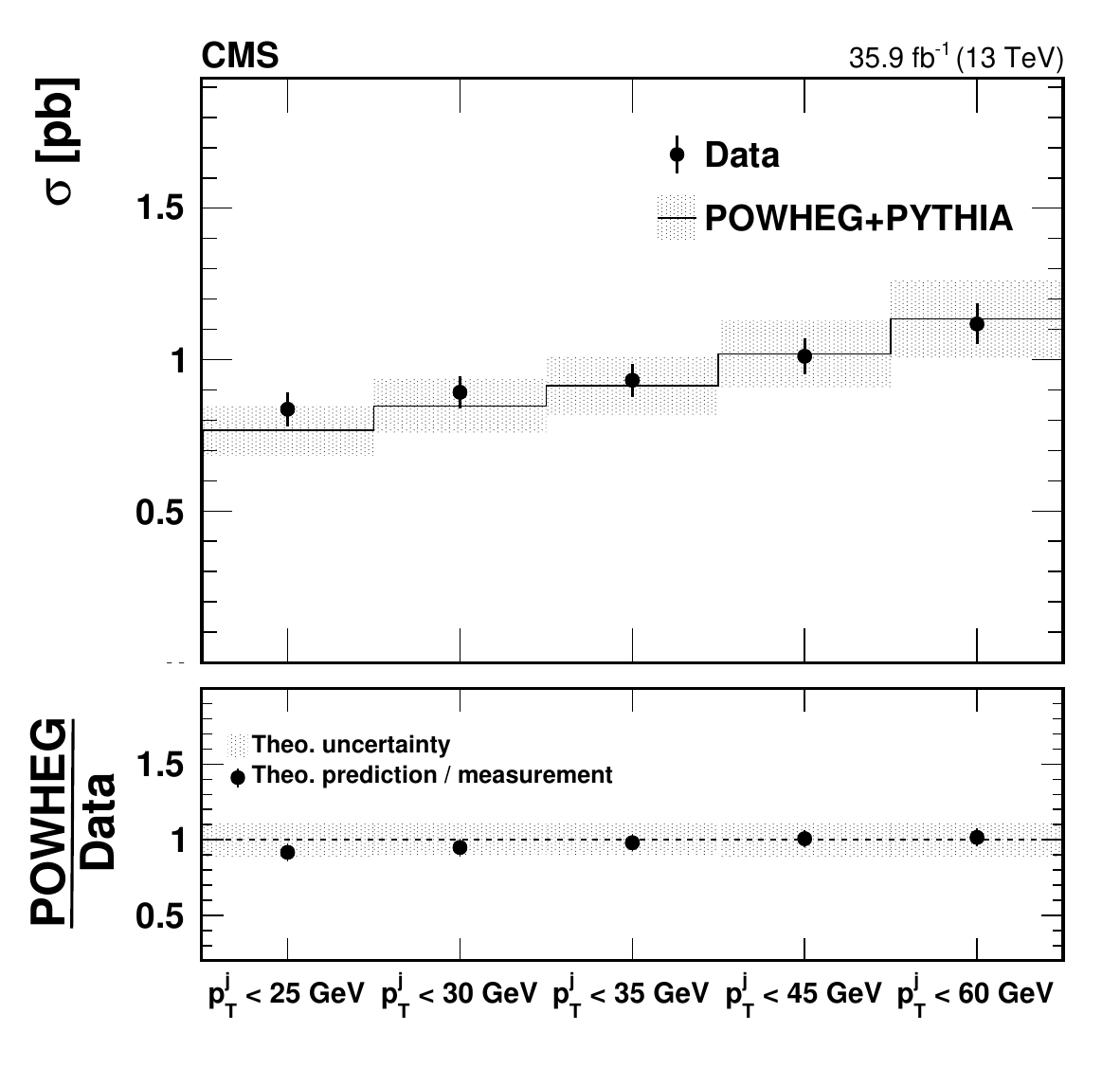}
\caption[.]{\label{fig:TAptthresholds}
The upper panel shows the fiducial cross sections for the production of $\WWpm+$\,0-jets
as the \pt threshold for jets is varied.
The fiducial region is defined by two opposite-sign leptons with $\pt>20\GeV$
and $\abs{\eta}<2.5$ excluding the products of $\tau$ lepton decay, and
$\mll>20\GeV$, $\qT>30\GeV$, and $\ptmiss>30\GeV$.  Jets must have
$\pt$ above the stated threshold, $\abs{\eta}<4.5$, and be separated from each of the two
leptons by $\Delta R > 0.4$.
The lower panel shows the ratio of the theoretical prediction to the measurement.
In both the upper and lower panels, the error bars on the data points represent the
total uncertainty of the measurement, and the
shaded band depicts the uncertainty of the MC prediction.
}
\end{figure}

The fiducial cross section for $\WWpm+$\,0-jets production  is also measured
as a function of the jet \pt threshold in the range 25--60\GeV
with the results listed in Table~\ref{tab:TAptthresholds} and displayed in Fig.~\ref{fig:TAptthresholds}.
The cross section is expected to increase with jet \pt threshold because the phase
space for zero jets increases.

\begin{table}[ht]
\centering
\topcaption[.]{\label{tab:TAptthresholds}
Fiducial cross section for the production of $\WWpm+$\,0-jets as the \pt threshold
for jets is varied.
The fiducial region is defined by two opposite-sign leptons with $\pt>20\GeV$
and $\abs{\eta}<2.5$ excluding the products of $\tau$ lepton decay, and
$\mll>20\GeV$, $\qT>30\GeV$, and $\ptmiss>30\GeV$.  Jets must have
$\pt$ above the stated threshold, $\abs{\eta}<4.5$, and be separated from each of the two
leptons by $\Delta R > 0.4$.
The total uncertainty is reported.}
\begin{scotch}{ c c c }
\pt threshold (GeV) & Signal strength & Cross section (pb) \\
\hline
 25 & $1.091 \pm 0.073$ & $0.836 \pm 0.056$ \\
 30 & $1.054 \pm 0.065$ & $0.892 \pm 0.055$ \\
 35 & $1.020 \pm 0.060$ & $0.932 \pm 0.055$ \\
 45 & $0.993 \pm 0.057$ & $1.011 \pm 0.058$ \\
 60 & $0.985 \pm 0.059$ & $1.118 \pm 0.067$ \\
\end{scotch}
\end{table}

\section{Normalized differential cross section measurements}\label{sec:diffWWxsec}
Differential cross sections are measured for the fiducial region defined above using
the sequential-cut, DF event selection.
The random forest selection is unsuitable for measuring these differential cross sections
because some of these kinematic quantities are used as inputs to the random forest classifiers.
These differential cross sections are normalized to the measured integrated fiducial cross section,
which for the DF final state (0- and 1-jet) is $0.782\pm0.053$\pb corresponding
to a signal strength of $1.022\pm0.069$.

For each differential cross section, a simultaneous fit to the reconstructed distribution 
is performed in the following manner.  An independent signal strength parameter is assigned
to each generator-level histogram bin. For the MC simulated events falling within a given
generator-level bin, a template histogram of the reconstructed kinematic quantity is formed.
The detector resolution is good for the quantities considered, so the template histogram
has a peak corresponding to the given generator-level bin; the contents of all bins below
and above the given generator-level bin are relatively low.  When the fit is performed,
the signal strengths are allowed to vary independently.  The correlations among bins
in the distribution of the reconstructed quantity are taken into account.
The fitted values of the signal strength parameters are applied to the generator-level
differential cross section to obtain the measured differential cross section.  

\begin{figure*}
\centering
\includegraphics[width=0.49\textwidth]{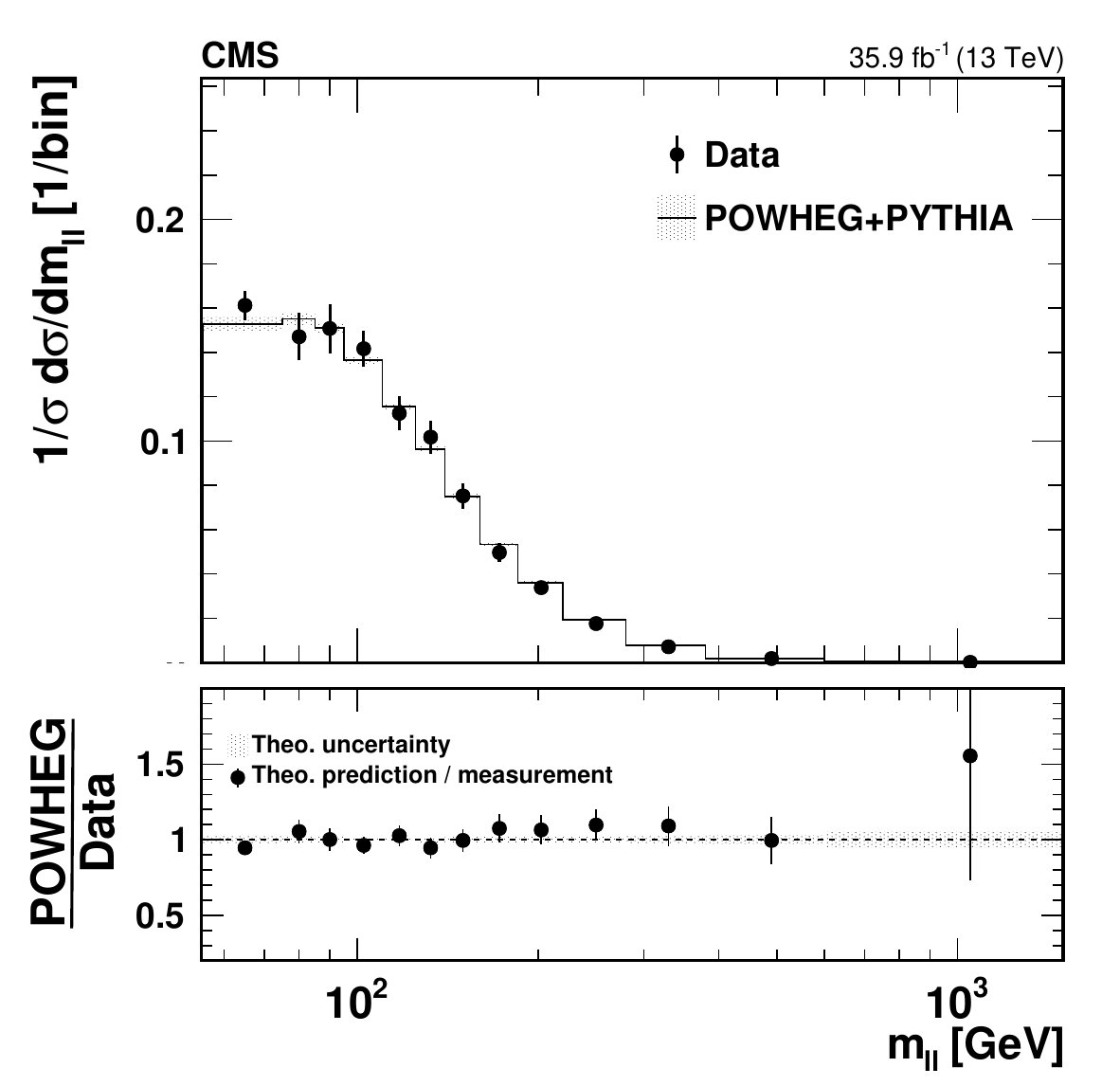}
\includegraphics[width=0.49\textwidth]{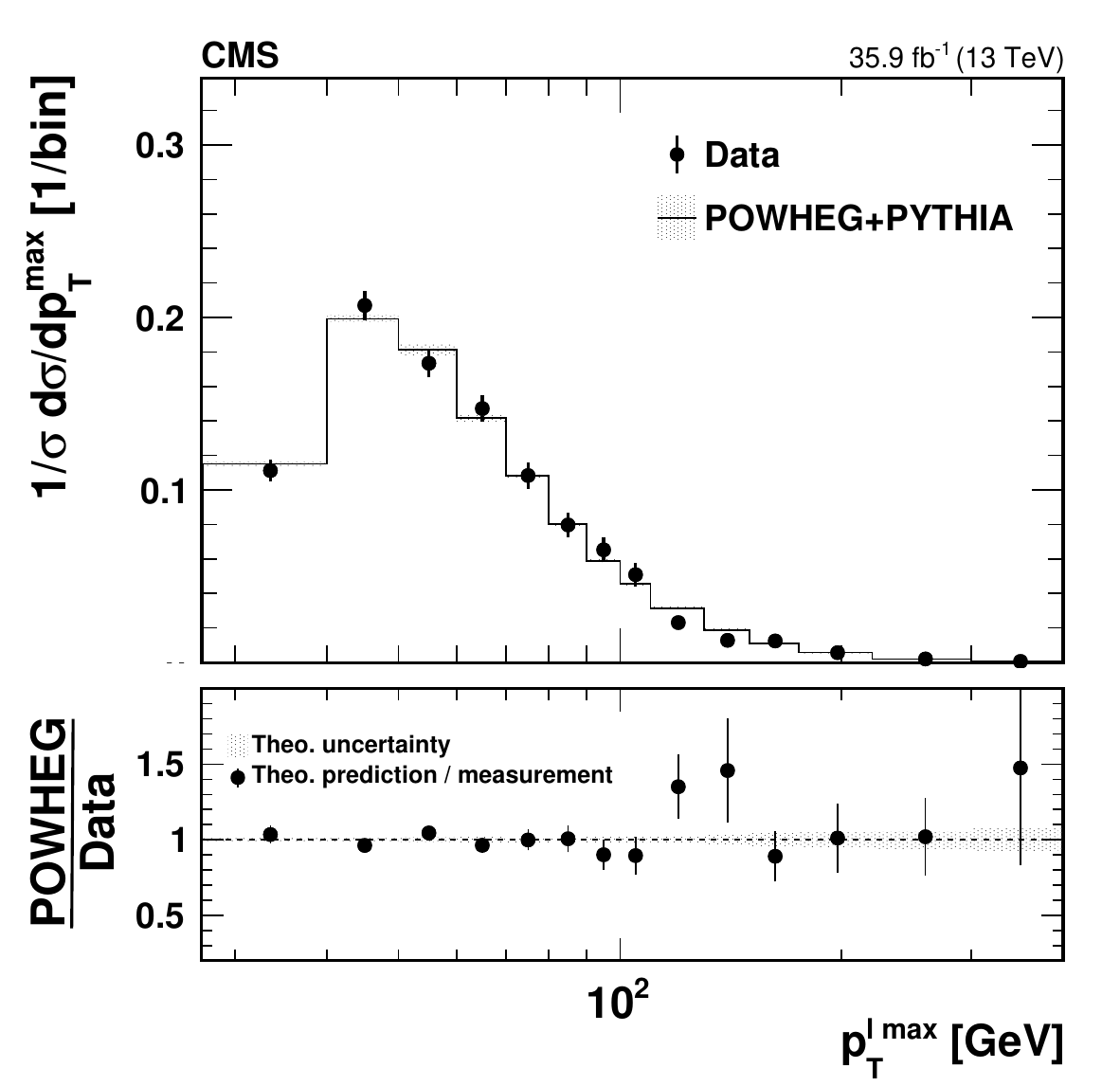}\\
\includegraphics[width=0.49\textwidth]{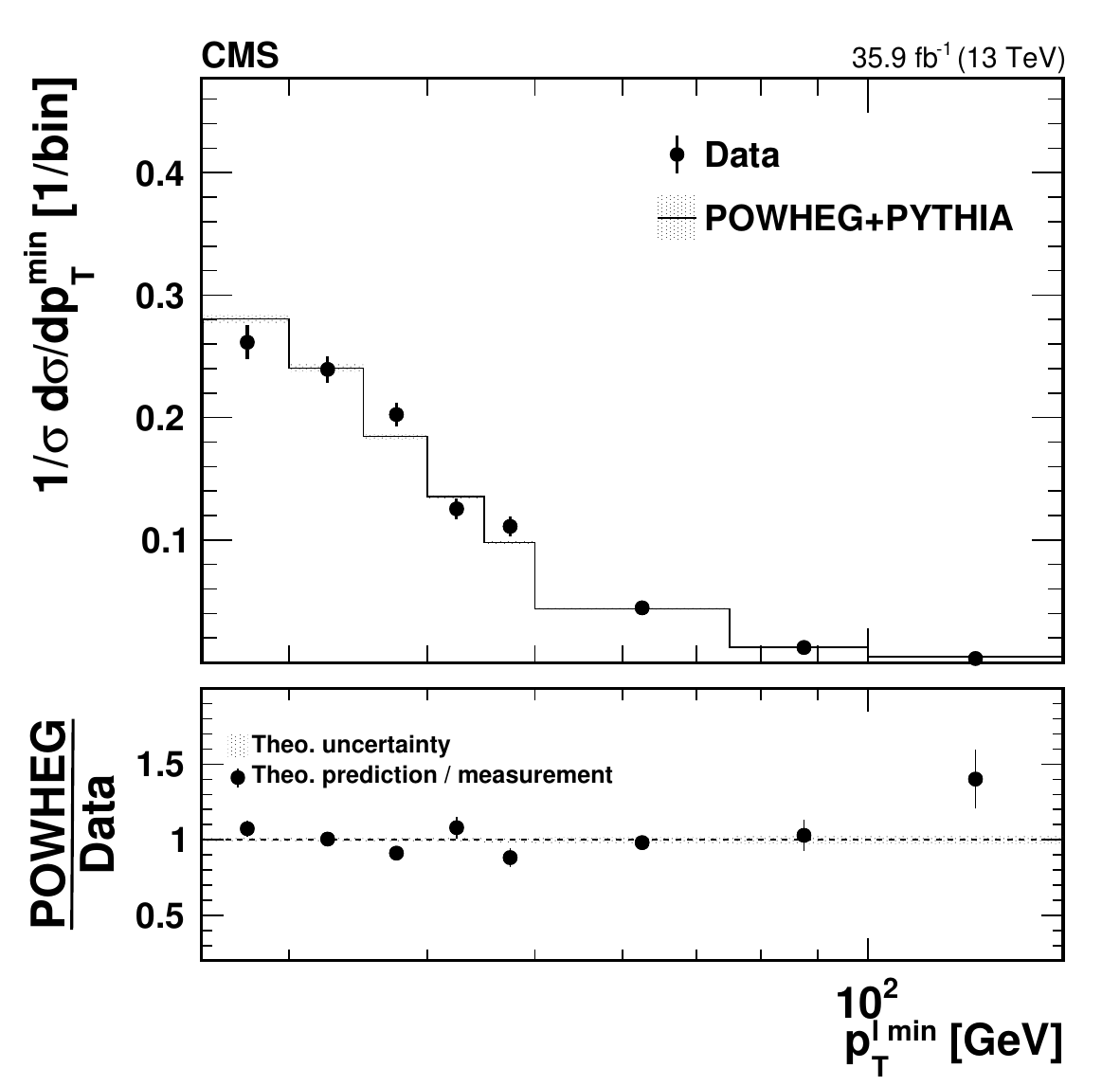}
\includegraphics[width=0.49\textwidth]{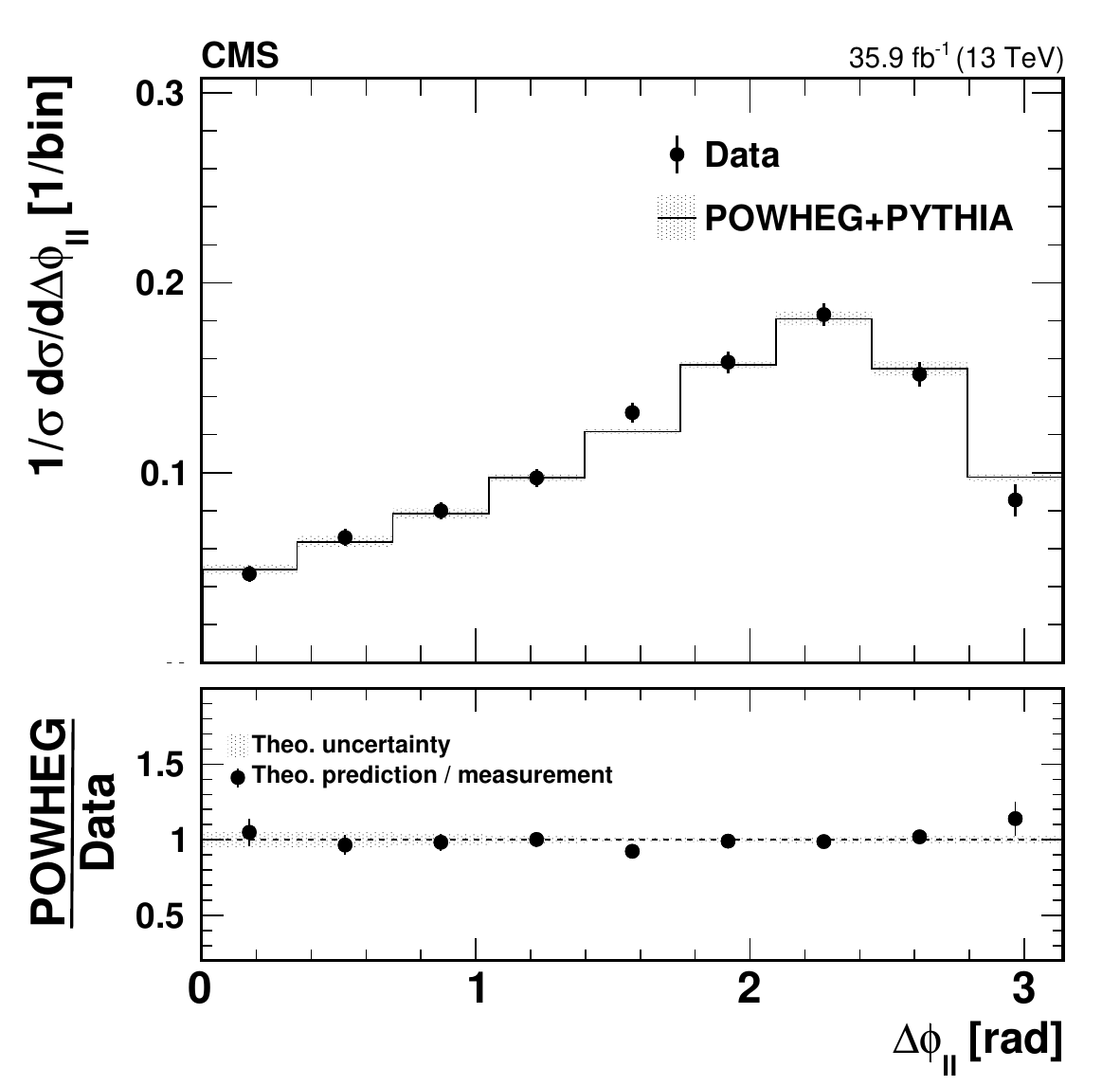}
\caption[.]{\label{fig:differential}
The upper panels show the 
normalized differential cross sections with respect to the dilepton mass \mll,
leading lepton \ptmax, trailing lepton \ptmin, and dilepton azimuthal angular separation $\delphill$,
compared to \POWHEG predictions.
The lower panels show the ratio of the theoretical predictions to the measured values.
The meaning of the error bars and the shaded bands is the same as in Fig.~\ref{fig:TAptthresholds}.}
\end{figure*}

Measurements of the differential cross sections with respect to the dilepton mass 
$(1/\sigma)\ddinline{\sigma}{\mll}$, the leading lepton transverse momentum 
$(1/\sigma)\ddinline{\sigma}{\ptmax}$, the trailing lepton transverse momentum
$(1/\sigma)\ddinline{\sigma}{\ptmin}$, and the angular separation between the leptons
$(1/\sigma)\ddinline{\sigma}{\delphill}$ are reported.
The measurements are compared to simulations generated with {\POWHEG}+\PYTHIA in Fig.~\ref{fig:differential}.

\section{Jet multiplicity measurement}\label{sec:NJets}
A measurement of the jet multiplicity tests the accuracy  of 
theoretical calculations and event generators.  
Signal \WWpm events are characterized by a low jet multiplicity
in contrast to \ttbar background events, which typically have two or
three jets.  The sequential event selection exploits this difference
by eliminating events with more than one jet and by separating 0-
and 1-jet event categories. The random forest selection, in contrast,
places no explicit requirements on the number of jets ($\NJ$) in an event,
and the separation of signal \WWpm events and \ttbar background
utilizes other event features listed in Table~\ref{tab:feature_list}.
As a consequence, a precise measurement of the fractions of events with
$\NJ=0$, 1, or $\ge2$ jets can be made.  For this measurement,
jets have $\pt>30\GeV$ and $\abs{\eta}<2.4$, and must be separated
from each of the selected leptons by $\Delta R > 0.4$.  The rejection of
events with one or more \PQb-tagged jets is still in effect;
however, the impact on the signal is very small.

The anti-\ttbar random forest produces a continuous score, $\STT$,
in the range  $0 \le \STT \le 1$, as explained in 
Section~\ref{sec:selectionRF}.   For the measurement of the jet
multiplicity presented in this section, the criterion against
\ttbar background is loosened to $\STTmin=0.2$ while 
$\SDYmin=0.96$ remains.  This looser requirement leads to a
signal efficiency for the random forest selection with a relatively
gentle variation with \NJ as shown in Table~\ref{tab:nj_eff}, and
also a more even variation of the efficiency as a function of \WWpt,
as shown in Fig.~\ref{fig:WWptcompare}.
These efficiencies are defined for the events passing the
random forest selection with respect to those passing the preselection  requirements.
The efficiency for the preselection is essentially independent of~\NJ.

\begin{table}[ht]
\centering
\topcaption[.]{\label{tab:nj_eff}
Efficiency for the random forest selection with respect to
preselected events as a function of jet multiplicity. 
The stated uncertainties are statistical only.}
\begin{scotch}{lccc}
Number of jets & 0 & 1 & $\ge 2$ \\
\hline
Efficiency     & $0.555 \pm 0.003$ & $0.448\pm 0.004$ & $0.290 \pm 0.004$ \\
\end{scotch}
\end{table}

Background contributions are subtracted from the observed numbers
of events as a function of \NJ and then corrections are applied for the
random forest efficiencies shown in Table~\ref{tab:nj_eff}.  The observed jet
multiplicity suffers from the migration of events from one \NJ bin to another due to
two experimental effects: first, pileup can produce extra jets (pileup),
and second, jet energy mismeasurements
can lead to jets with true \pt below the 30\GeV threshold being
accepted and others with true \pt above 30\GeV being rejected.
Pileup jets only increase the number of jets in an event while energy calibration
and resolution leads to both increases and decreases in \NJ.  Because of the 
falling jet \pt distribution, the jet energy resolution leads to increases
in \NJ more often than to decreases.

The two sources of event migration are corrected in two distinct steps.
The signal MC event sample is used to build two response matrices:
$\RPU$ for pileup and $\RJER$ for detector effects, in particular, jet energy
resolution.  The reconstructed jet multiplicity for the signal process is
given by $\vec{v} = \RPU\,\RJER\,\vec{t}$ where $\vec{v}$ and $\vec{t}$
are vectors representing the multiplicity distribution;
$\vec{t}$ represents the MC ``truth'' as inferred from 
generator-level jets and $\vec{v}$ is the reconstructed distribution.
Generator-level jets are reconstructed from
generated stable particles, excluding neutrinos, with the clustering algorithm used to 
reconstruct jets in data.  These jets must satisfy $\pt>30\GeV$ and
$\abs{\eta}<2.4$ and must be separated by $\Delta R > 0.4$ from both of the
two leptons from \PW boson decays.  Reconstructed and generator-level jets
are said to match if they have $\Delta R < 0.4$.  On the basis of the
simulated signal event sample, the two response matrices are close
to being diagonal:
\begin{linenomath}
\begin{equation*}
\begin{aligned}
  \RPU =&
    \begin{pmatrix}
       0.986 & 0 & 0 \\
       0.013 & 0.985 & 0 \\
       0.001 & 0.015 & 1 \\
    \end{pmatrix} 
 \\[3\cmsTabSkip]
  \RJER = &
    \begin{pmatrix}
       0.963 & 0.060 & 0.003 \\
       0.036 & 0.891 & 0.090 \\
       0.001 & 0.049 & 0.906 \\
    \end{pmatrix}.
\end{aligned}
\end{equation*}
\end{linenomath}
Here, the columns correspond to $\NJ=0$, 1, $\ge2$ for generator-level
jets, and the rows to the same for reconstructed jets.

The response matrices are used to unfold the distribution of jet multiplicities
according to $\vec{u} = \RJERI\,\RPUI\,\vec{v}$.  
No regularization procedure is applied.   The fractions of events with 
$\NJ=0$, 1, $\ge2$ jets are obtained by normalizing $\vec{u}$ to unit norm: 
the unfolded result is $\vec{w} = \vec{u} / \abs{\vec{u}}$.

All systematic uncertainties are reevaluated for the jet multiplicity
measurement.  Since the observables are essentially yields normalized to the total number of events,
systematic uncertainties from the integrated luminosity and lepton efficiency
are negligible.  The statistical uncertainty in the response matrix
is also negligible.  Nonnegligible uncertainties are obtained for
the jet energy scale and resolution, for pileup reweighting, and for
reweighting of the \WWpt spectrum.  The total relative uncertainties for 
the elements of the response matrix are:
\begin{linenomath}
\begin{equation*}
 \begin{pmatrix}
     0.011 & 0.193 & 0.374 \\
     0.210 & 0.007 & 0.140 \\
     0.305 & 0.181 & 0.015 \\
 \end{pmatrix}.
\end{equation*}
\end{linenomath}
Although the relative uncertainty of the off-diagonal matrix elements
is large, those elements themselves are small, so a precise measurement
is still achievable.

\begin{figure}[htb]
\centering
\includegraphics[width=0.48\textwidth]{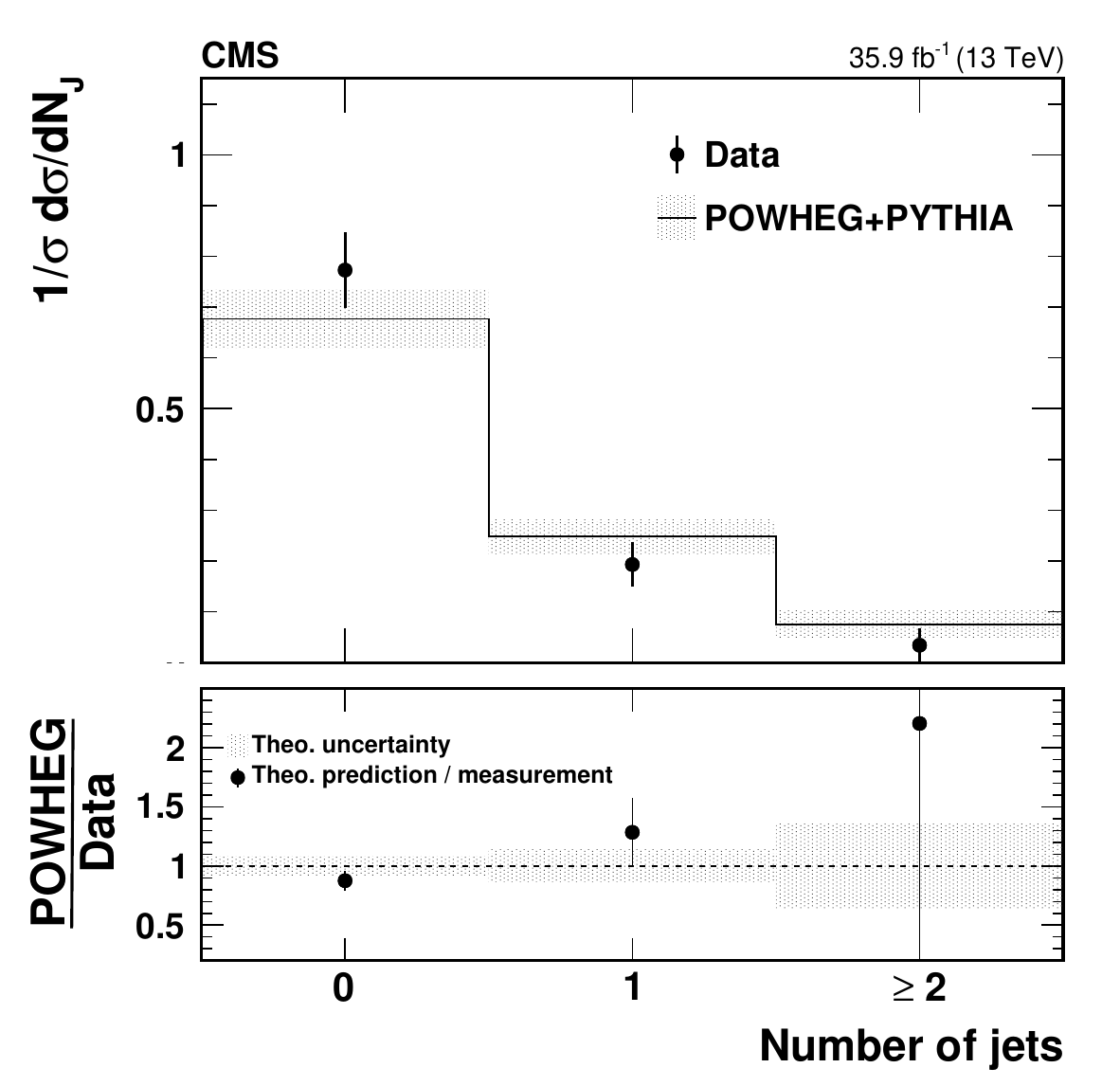}
\caption[.]{\label{fig:NJ} The upper panel shows the
fractions of events with $\NJ=0$, 1, $\ge2$ jets.  
The filled circles represent the data after backgrounds are subtracted
and pileup and energy resolution are taken into account.
The solid lines represent the {\POWHEG}+\PYTHIA prediction.
The lower panel shows the ratio  of the theoretical prediction to the measurement.
The meaning of the error bars and the shaded bands is the same as in Fig.~\ref{fig:TAptthresholds}.}
\end{figure}

Table~\ref{tab:NJ} reports the measured fractions of events with \NJ jets.
The fractions before unfolding for pileup and jet energy resolution are
listed, as well as the prediction based on \POWHEG weighted
to correct the \WWpm\,\pt spectrum.   Figure~\ref{fig:NJ} shows a comparison
of the measured fractions and the prediction from \POWHEG.  For this
prediction, the \WWpt spectrum is reweighted as described in
Section~\ref{sec:unctheory}.

\begin{table*}[ht]
\centering
\topcaption[.]{\label{tab:NJ}
Fractions of events with $\NJ=0$, 1, $\ge2$ jets.
The first uncertainty is statistical and the second combines systematic
uncertainties from the response matrix and from the background subtraction.
}
\begin{scotch}{ l  c c c }
Number of jets & 0 & 1 & $\ge 2$ \\
\hline
Before unfolding        & $0.795\pm0.007\pm0.053$ & $0.180\pm0.006\pm0.039$ & $0.025\pm0.005\pm0.018$ \\
After unfolding         & $0.773\pm0.008\pm0.075$ & $0.193\pm0.007\pm0.043$ & $0.034\pm0.006\pm0.033$ \\
Predicted               & $0.677\pm0.007\pm0.058$ & $0.248\pm0.007\pm0.033$ & $0.075\pm0.006\pm0.026$ \\
\end{scotch}
\end{table*}

\section{Limits on dimension-6 Wilson coefficients}\label{sec:aTGC}
In the framework of effective field theory, new physics can be 
described in terms of an infinite series of new interaction terms organized as an 
expansion in the mass dimension of the corresponding operators~\cite{Buchmuller:1985}.
The dimension-4 operators of the SM comprise the zeroth term of the expansion.
The series can be understood as coming from the integration of heavy fields 
in an ultraviolet-complete theory, which itself is renormalizable and unitary. 
When testing for the presence of these higher-dimensional operators,
it is assumed that just one or two operators have nonvanishing coefficients
in order to reduce the computational burden.
A truncated series, \eg, a series including the SM and dimension-6 operators 
only, is not renormalizable and will violate tree-level unitarity at some energy
scale.  Consequently, the truncated series is useful only when the 
scale of new physics is large compared to the energies accessible in the
given final state, in which case terms including higher-dimensional
operators are suppressed.

In the electroweak sector of the SM, the first higher-dimensional 
operators containing only massive boson fields are
dimension-6~\cite{Grzadkowski:2010es,Degrande:2012wf}:
\begin{linenomath}
\begin{equation*}
\nonumber
\begin{split}
\mathcal{O}_{\PW\PW\PW} = \frac{c_{\PW\PW\PW}}{\Lambda^2}W_{\mu\nu}W^{\nu\rho}W_{\rho}^{~~\mu}\\
\mathcal{O}_{\PW} = \frac{c_{\PW}}{\Lambda^2} (D^{\mu}\Phi)^{\dagger} W_{\mu\nu} (D^{\nu}\Phi)\\
\mathcal{O}_{B} = \frac{c_{B}}{\Lambda^2} (D^{\mu}\Phi)^{\dagger} B_{\mu\nu} (D^{\nu}\Phi)\\
\widetilde{\mathcal{O}}_{\PW\PW\PW} = \frac{\widetilde{c}_{\PW\PW\PW}}{\Lambda^2}\widetilde{W}_{\mu\nu}W^{\nu\rho}W_{\rho}^{~~\mu}\\
\widetilde{\mathcal{O}}_{\PW} = \frac{\widetilde{c}_{\PW}}{\Lambda^2} (D^{\mu}\Phi)^{\dagger} \widetilde{W}_{\mu\nu} (D^{\nu}\Phi).\\
\end{split}
\end{equation*}
\end{linenomath}
The  gauge group indices are suppressed for clarity and the
mass scale $\Lambda$ has been factored out from the Wilson coefficients $c$ and $\widetilde{c}$.
The tensor $W_{\mu\nu}$ is the $SU(2)$ field strength, $B_{\mu\nu}$ is the $U(1)$ field strength,
$\Phi$ is the Higgs doublet, and operators with a tilde are the magnetic duals of the 
field strengths. The first three operators are CP conserving, while the last two are not.
In this analysis, only the CP conserving operators are considered.

\begin{figure}[htpb]
\centering
\includegraphics[width=0.4\textwidth]{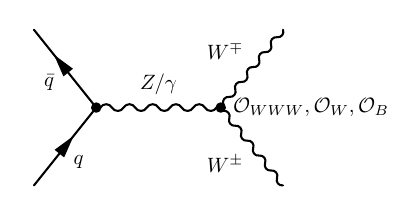}
\caption{\label{atgc:fig1}One of the Feynman diagrams through which dimension-6
operators modify the $\Pp\Pp\to\WWpm$ cross section.}
\end{figure}

These operators contribute to several multiboson scattering processes at tree level.
The operator $\mathcal{O}_{\PW\PW\PW}$ modifies vertices with three to six vector bosons,
while $\mathcal{O}_{\PW}$ and $\mathcal{O}_{B}$ modify both ${\PH}\PV\PV$ vertices and vertices
with three or four vector bosons. The focus in this analysis is on modifications to the
vertices $\PH\PWp\PWm$, $\gamma\WWpm$, and $\PZ\WWpm$ since they lead to deviations
of the $\Pp\Pp\to\WWpm$ cross section via diagrams of the kind shown in Fig.~\ref{atgc:fig1}.

The analysis is based on the DF event sample selected in
the sequential cut analysis.  The SF event sample is not used because
the contamination from Drell--Yan processes is larger and the selected event
sample itself is smaller.  The 0- and 1-jet categories are analyzed separately.
The signal region and the top quark control region are both included in the
analysis.

The invariant mass \memu distribution is used to test for dimension-6
operators.  The quantity \memu is well measured and is not sensitive to higher-order QCD effects
and jet energy calibration issues.  Furthermore, the \memu distribution is more sensitive
to higher-dimensional operators than other observables based on lepton kinematics.
In order to suppress the Higgs boson contribution and enhance the sensitivity
to higher-dimensional operators, the requirement $\memu>100\GeV$ is imposed.
The remaining Higgs boson contributions are considered part of the signal.
Variations of the relatively small $\PV\PZ$ background processes 
due to dimension-6 operators are neglected.

The measurement of the Wilson coefficients uses templates of \memu with
the following bins:
$[100$, 200, 300, 400, 500, 600, 700, 750, 800, 850, 1000, $\infty]\GeV$;
the last bin contains all events with $\memu>1\TeV$. 
This choice minimizes the expected 95\% confidence level (\CL) intervals for all coefficients
(with fixed mass scale  $\Lambda$) while populating each bin adequately.
The highest bin is the one with the greatest statistical power largely because of
the presence of multiple momentum factors in the Feynman diagrams associated 
with the higher-dimensional operators (Fig.~\ref{atgc:fig1}).

In order to construct the \memu templates, the weights calculated for each event
are used to build a parametrized model of the expected yield in each bin as a
function of the coefficients (with fixed $\Lambda$).
More precisely, for each bin, a fit is performed of a second-order polynomial
to the ratios of the expected signal yield with nonzero coefficients to the
one without (SM).   When only one coefficient is taken to be nonzero, then
the fit is performed to five points, and when two coefficients are taken to
be nonzero, the fit is performed to a $5\times5$ grid. These fits are carried
out for the 0- and 1-jet categories separately.

A binned maximum likelihood fit of the \memu templates to the data is carried out.
The likelihood is computed using the Poisson probability for each bin $i$ with 
$\Niexp$ expected events and $\Niobs$ observed events. 
Each source of uncertainty is modeled with a log-normal distribution
$\pi_{ij}(\theta_{j})$ where $\theta_{j}$ is the nuisance parameter
for a source of uncertainty $j$ as discussed in Section~\ref{sec:systematics}.
The expected yields $\Niexp$ are functions of the nuisance parameters $\theta_{j}$.
The likelihood is computed from the product over all bins~$i$:
\begin{linenomath}
\begin{equation}
\nonumber
 L = \prod_{i=i}^{N} \left[\re^{-\Niexp}(\Niexp)^{\Niobs} \, \prod_{j=i}^{M} \pi_{ij}({\theta}_{j})\right]  
\end{equation}
\end{linenomath}
where the $N!$ term has been neglected.  The nuisance parameters for the
systematic uncertainties are profiled for each dimension-6 operators hypothesis. 

\begin{figure}
\centering
\includegraphics[width=\cmsFigWidthTwo]{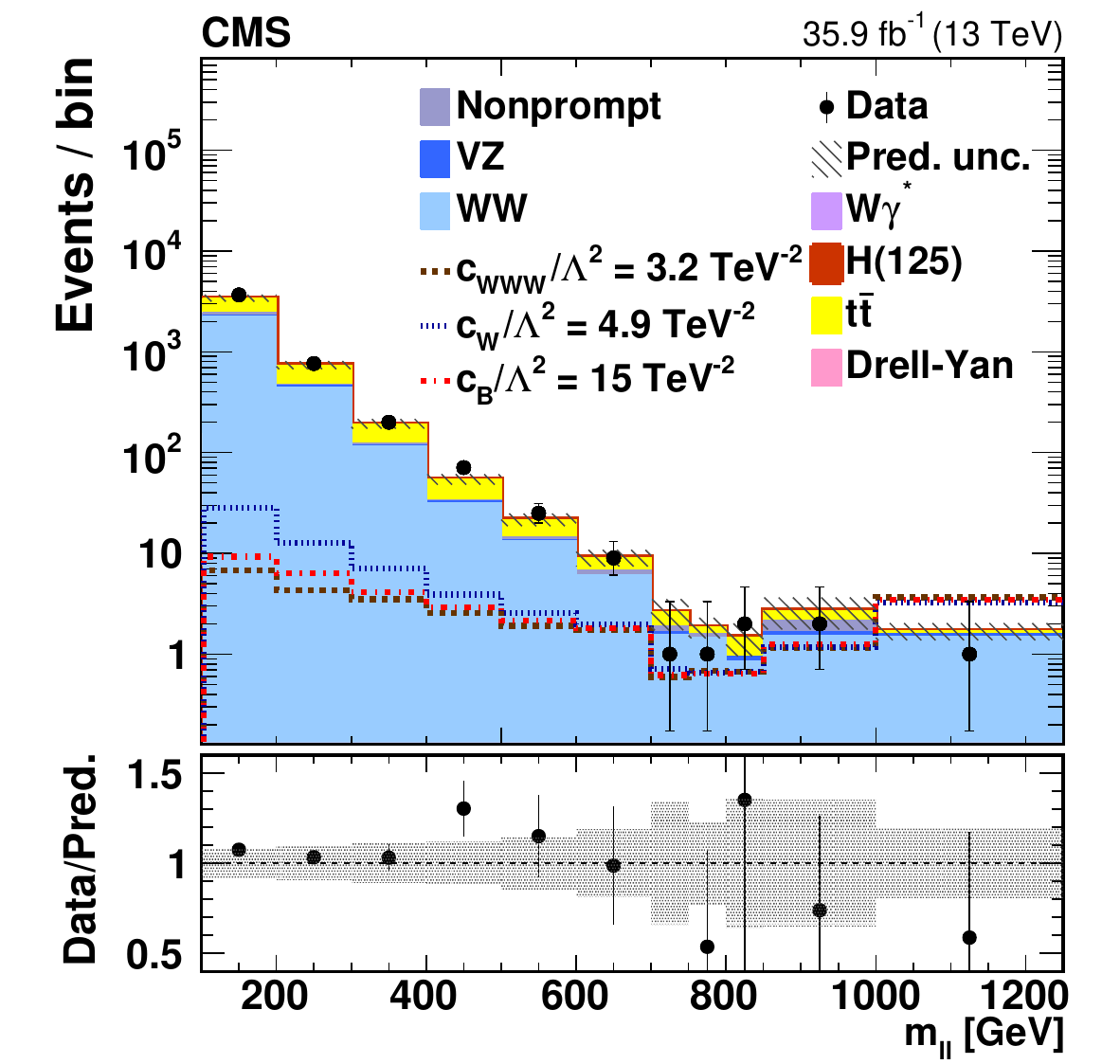}
\includegraphics[width=\cmsFigWidthTwo]{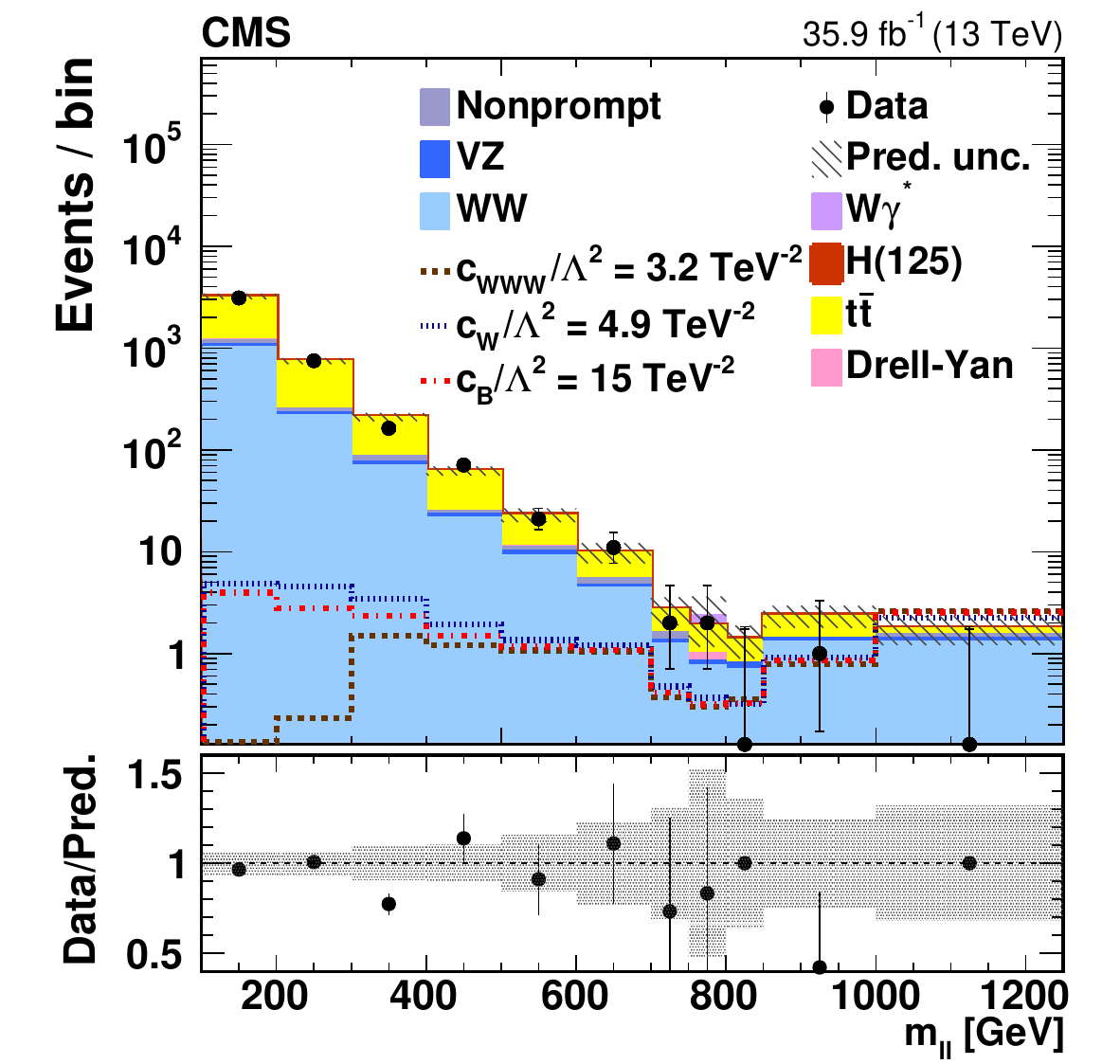}
\caption{\label{atgc:mll_ALL}
  Comparison of the template fits to the observed \memu distributions in the 0-jet (left)
  and 1-jet (right) categories.  The non-SM contributions for $c_{\PW\PW\PW}/\Lambda^2=3.2\TeV^{-2}$,
  $c_{\PW}/\Lambda^2=4.9\TeV^{-2}$, and $c_{B}/\Lambda^2=15.0\TeV^{-2}$ are shown, not stacked
  on top of the other contributions.
  In the plot on the right, the decrease in the non-SM contribution at low \memu is not 
  statistically significant and results from
  limited precision in the subtraction of two large yields (SM and SM+non-SM).
  The last bin contains all events with reconstructed $\memu>1\TeV$.
  The error bars on the data points represent the statistical uncertainties for the data, and
  the hatched areas represent the total uncertainty for the predicted yield in each bin.  }
\end{figure}

Figure~\ref{atgc:mll_ALL} shows the results of the template fits to
the observed \memu distributions.  The expected signal distributions 
for three values of the coefficients close to the 95\% \CL expected limits on those
coefficients are also shown (not stacked); the largest impact of nonzero coefficients
is seen for $\memu>850\GeV$.

\begin{figure}
\centering
\includegraphics[width=\cmsFigWidthFix]{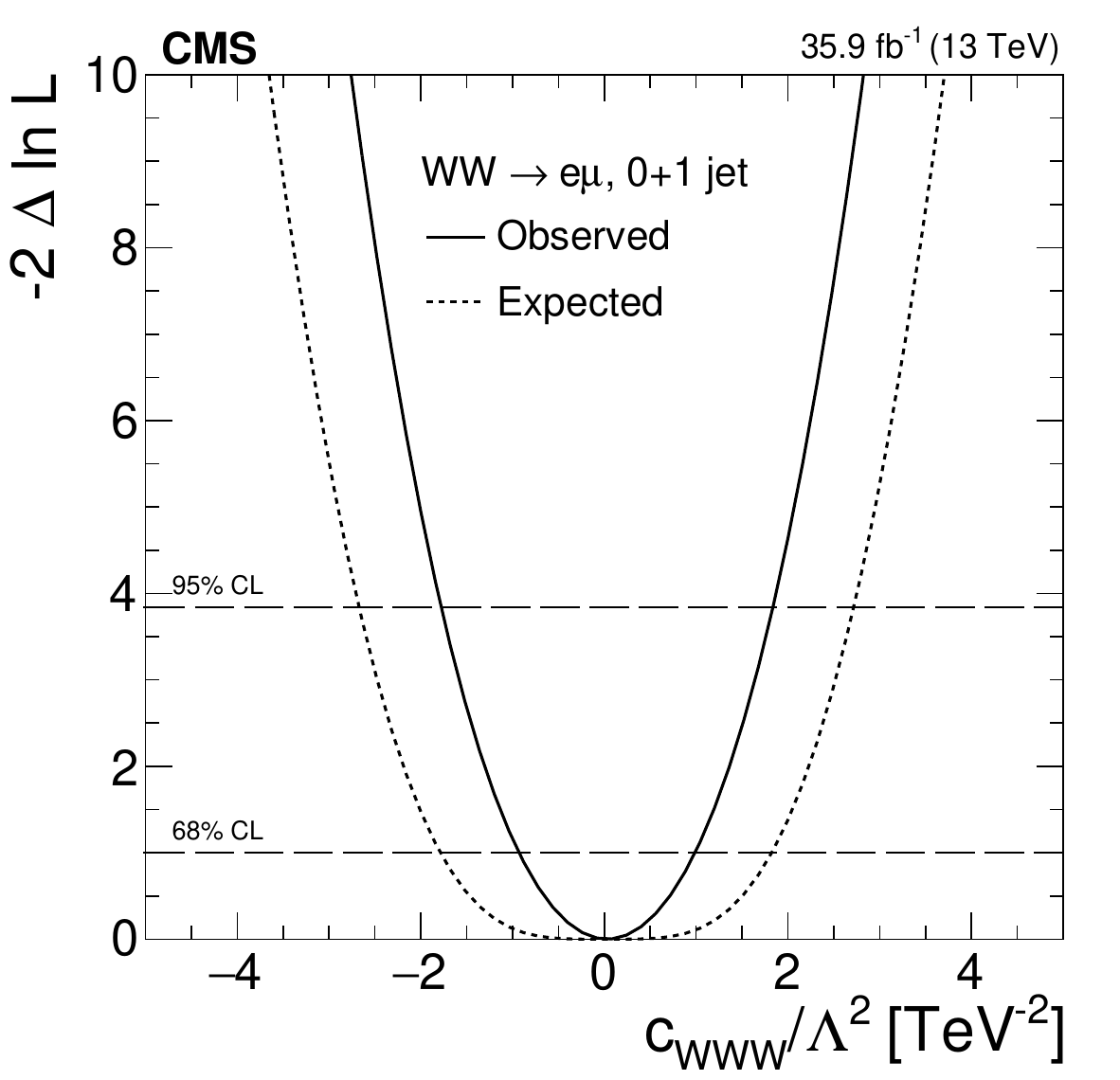}
\includegraphics[width=\cmsFigWidthFix]{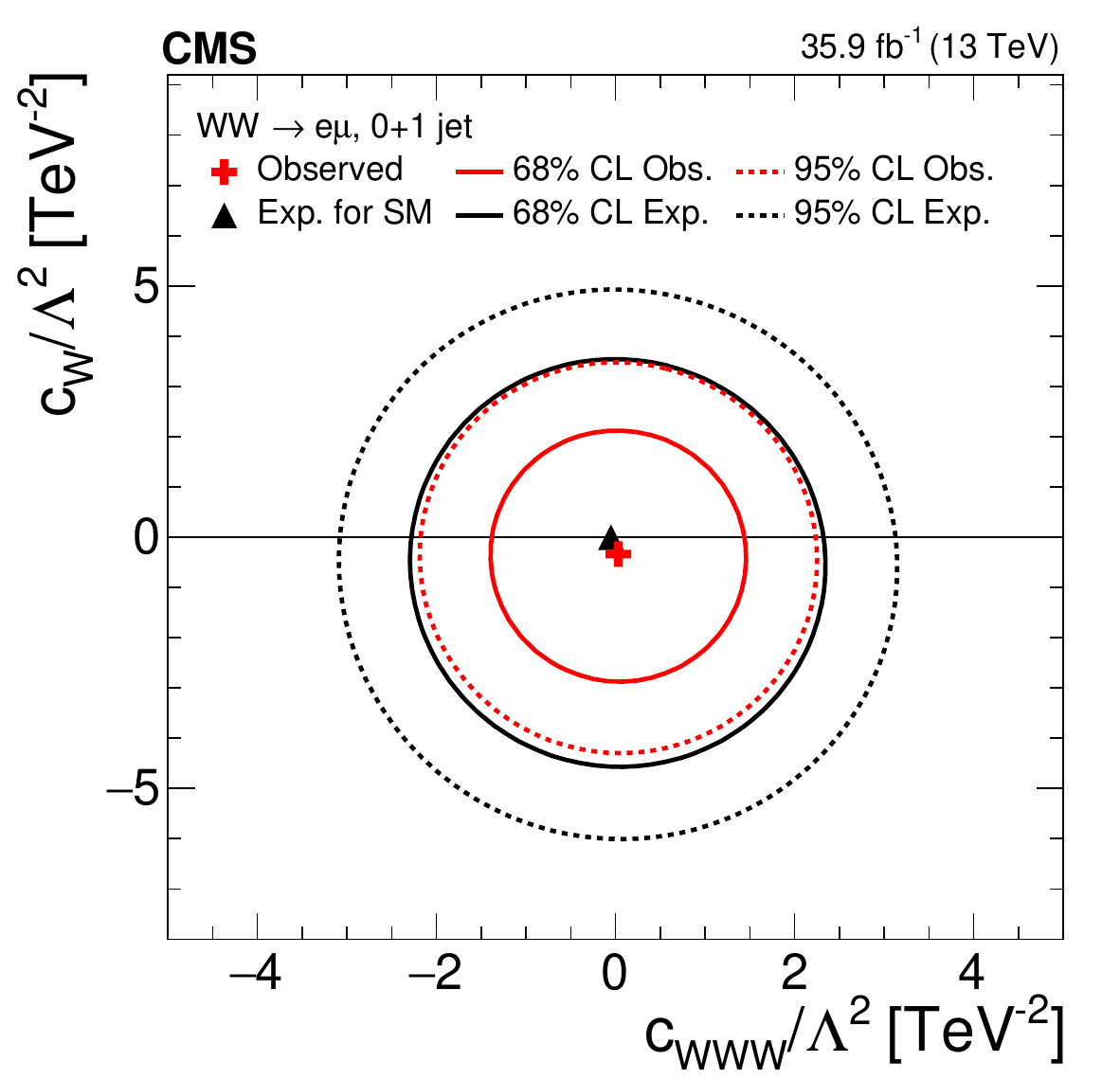}\\
\includegraphics[width=\cmsFigWidthFix]{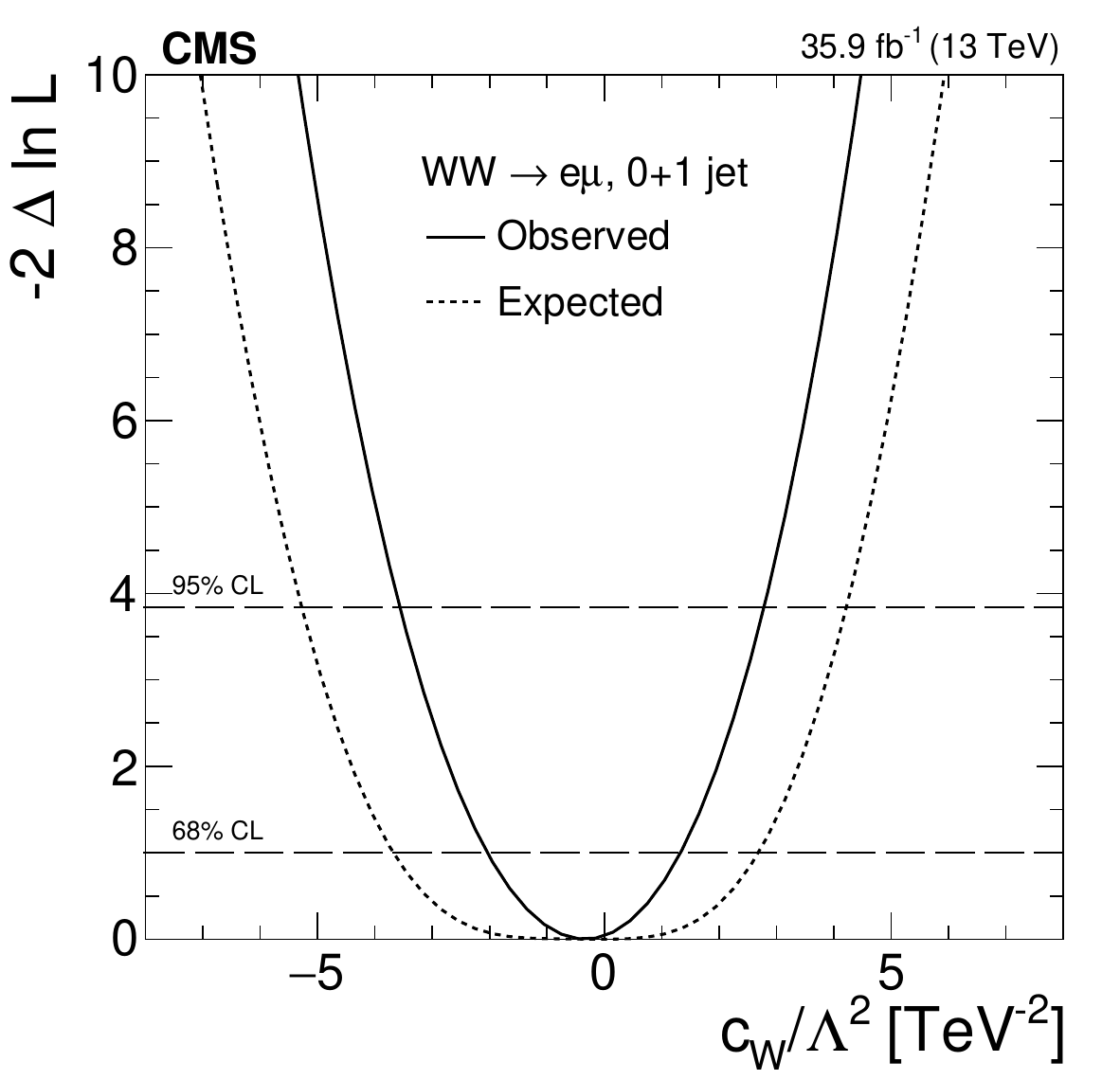}
\includegraphics[width=\cmsFigWidthFix]{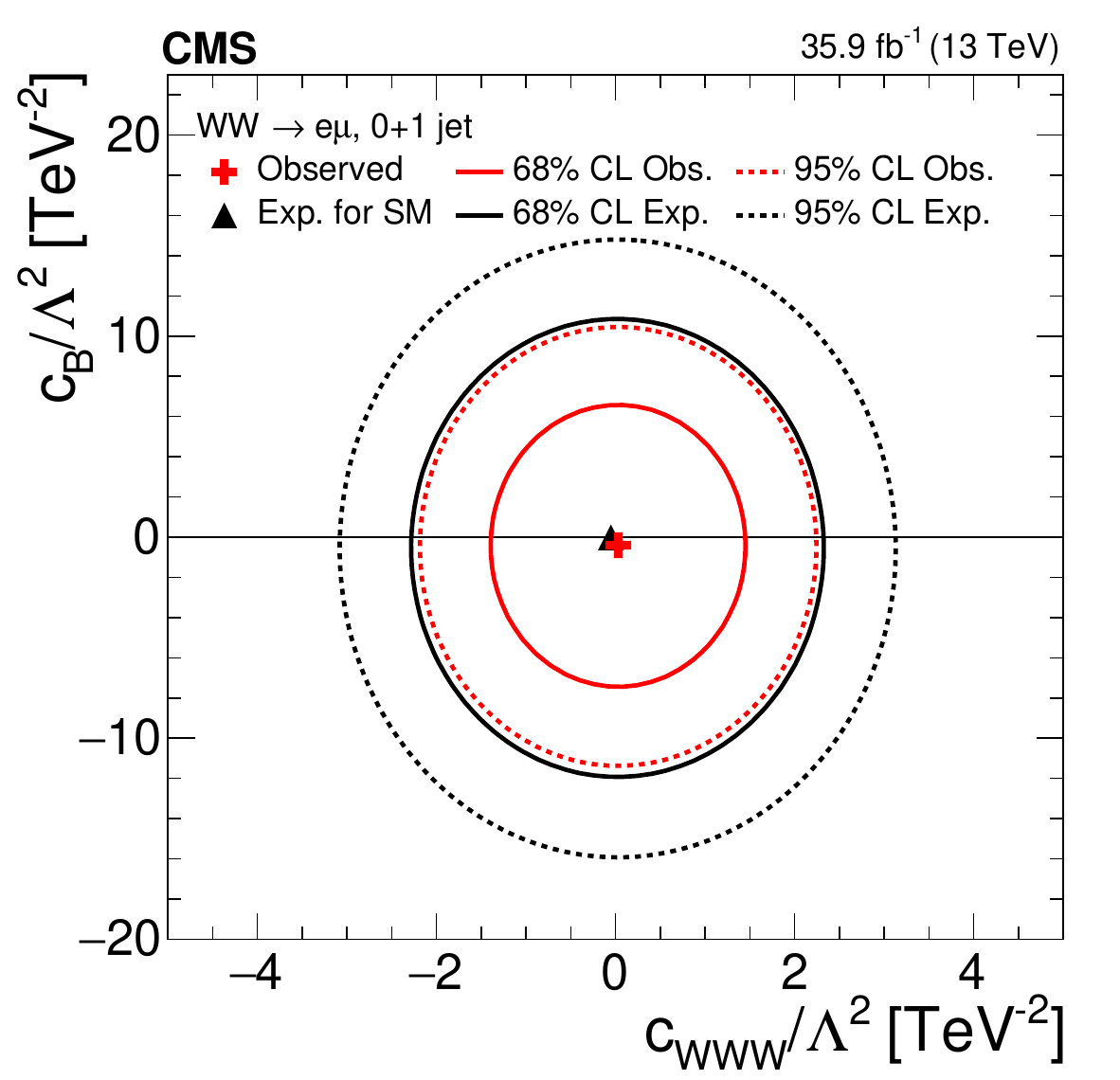}\\
\includegraphics[width=\cmsFigWidthFix]{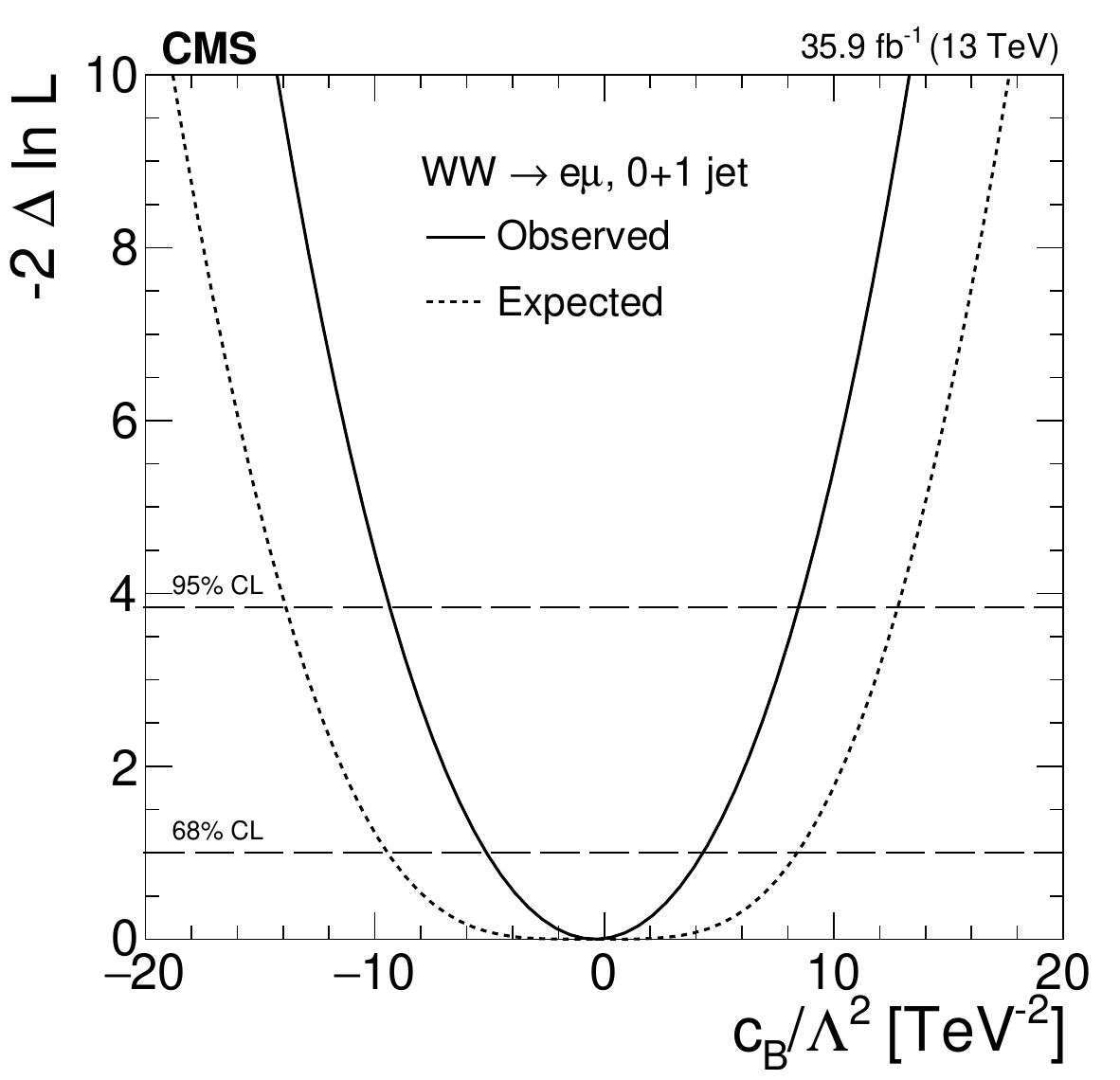}
\includegraphics[width=\cmsFigWidthFix]{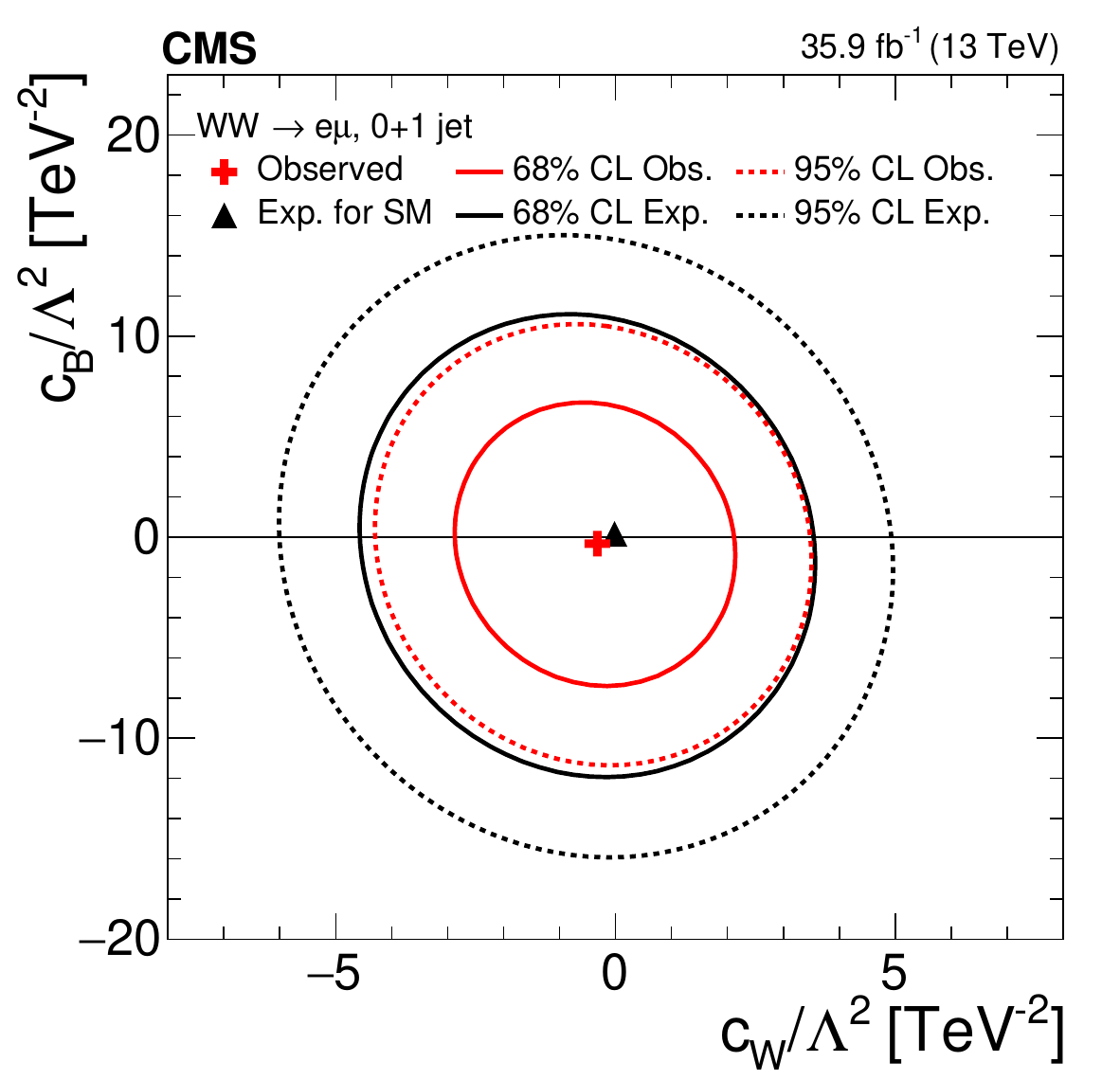}
\caption{\label{atgc:1DLim2DLim}
  On the left,
  the expected and observed $-2\Delta \ln L$ curves for the $c_{\PW\PW\PW}/\Lambda^2$, $
  c_{\PW}/\Lambda^2$, and $c_{B}/\Lambda^2$ combining the 0- and 1-jet categories.
  On the right,
  the expected and observed 68 and 95\% confidence level contours in the $(c_{\PW\PW\PW}/\Lambda^2, c_{\PW}/\Lambda^2)$,
  $(c_{\PW\PW\PW}/\Lambda^2, c_{B}/\Lambda^2)$, and $(c_{\PW}/\Lambda^2, c_{B}/\Lambda^2)$ planes
  combining the 0- and 1-jet categories.}
\end{figure}

Figure~\ref{atgc:1DLim2DLim} (left) shows the curves of $-2\Delta \ln L = -2( \ln L - \ln L_{\mathrm{min}} )$
for the three dimension-6 operators considered here; the 0- and 1-jet categories have been combined.
The corresponding 68 and 95\% \CL intervals are reported in Table~\ref{atgc:tbl3}.
The observed limits are stronger than expected due to a deficit of events at high \memu.
In all cases, they are within two standard deviations of the expected limits as determined by pseudo-experiments.
The observed limits are about a factor of two more stringent than recent results
reported by the ATLAS Collaboration~\cite{ATLAS:2019nkz} and the previous \WWpm results
from the CMS Collaboration~\cite{Khachatryan:2015sga}.  The sensitivity of this analysis to
$c_{\PW\PW\PW}$ and $c_{\PW}$ is similar to the CMS $\PW\PZ$ analysis~\cite{Sirunyan:2019bez} and
is much better for $c_{B}$.  Finally, the sensitivity is slightly weaker than for the CMS analysis
of \WWpm and $\PW\PZ$ production in lepton and jets events~\cite{Sirunyan:2019gkh}.
Figure~\ref{atgc:1DLim2DLim} (right) shows the expected and observed 68 and 95\% confidence level contours
for pairs of Wilson coefficients.

\begin{table}
\centering
\topcaption{\label{atgc:tbl3}
Expected and observed 68 and 95\% confidence intervals on the measurement of the Wilson coefficients
associated with the three CP-conserving, dimension-6 operators.} 
\begin{scotch}{c cc cc}
Coefficients & \multicolumn{2}{c}{68\% confidence interval} & \multicolumn{2}{c}{95\% confidence interval} \\
 ($\TeV^{-2}$) & expected & observed & expected & observed \\
\hline
 & & & &  \\ [-2ex]
$c_{\PW\PW\PW}/\Lambda^{2}$ & $[-1.8, 1.8]$ & $[-0.93, 0.99]$ & $[-2.7, 2.7]$ & $[-1.8, 1.8]$ \\
$c_{\PW}/\Lambda^{2}$      & $[-3.7, 2.7]$ & $[-2.0, 1.3]$ & $[-5.3, 4.2]$ & $[-3.6, 2.8]$ \\
$c_{B}/\Lambda^{2}$        & $[-9.4, 8.4]$ & $[-5.1, 4.3]$ & $[-14, 13]$ & $[-9.4, 8.5]$ \\
\end{scotch}
\end{table}

\section{Summary}\label{sec:conclusions}
Measurements of \WWpm boson pair production in proton-proton 
collisions at $\sqrt{s} = 13\TeV$ was performed. The analysis is based on 
data collected with the CMS detector at the LHC corresponding to an integrated luminosity
of 35.9\fbinv. Candidate events were selected that have two leptons (electrons or muons) with
opposite charges.  Two analysis methods were described.  The first
method imposes a sequence of requirements on kinematic quantities to suppress
backgrounds, while the second uses a pair of random forest classifiers.
The total production cross section is
$\sigmatotTA = 117.6 \pm 1.4\stat \pm 5.5\syst \pm 1.9\thy \pm 3.2\lum\pb = 117.6 \pm 6.8\pb$,
where the individual uncertainties are statistical, experimental systematic,
theoretical, and of integrated luminosity;
this measured value is consistent with the next-to-next-to-leading-order theoretical prediction $\sigmaNNLOval$.
Fiducial cross sections are also measured including the change in the
0-jet fiducial cross section with jet transverse momentum threshold.
Normalized differential cross sections are measured and compared with next-to-leading-order SM predictions.
Good agreement is observed.
The normalized jet multiplicity distribution in \WWpm events is measured.
Finally, bounds on coefficients of dimension-6 operators in the context of
an effective field theory are set using electron-muon invariant mass
distributions.

\begin{acknowledgments}
  We congratulate our colleagues in the CERN accelerator departments for the excellent performance of the LHC and thank the technical and administrative staffs at CERN and at other CMS institutes for their contributions to the success of the CMS effort. In addition, we gratefully acknowledge the computing centers and personnel of the Worldwide LHC Computing Grid for delivering so effectively the computing infrastructure essential to our analyses. Finally, we acknowledge the enduring support for the construction and operation of the LHC and the CMS detector provided by the following funding agencies: BMBWF and FWF (Austria); FNRS and FWO (Belgium); CNPq, CAPES, FAPERJ, FAPERGS, and FAPESP (Brazil); MES (Bulgaria); CERN; CAS, MoST, and NSFC (China); COLCIENCIAS (Colombia); MSES and CSF (Croatia); RIF (Cyprus); SENESCYT (Ecuador); MoER, ERC IUT, PUT and ERDF (Estonia); Academy of Finland, MEC, and HIP (Finland); CEA and CNRS/IN2P3 (France); BMBF, DFG, and HGF (Germany); GSRT (Greece); NKFIA (Hungary); DAE and DST (India); IPM (Iran); SFI (Ireland); INFN (Italy); MSIP and NRF (Republic of Korea); MES (Latvia); LAS (Lithuania); MOE and UM (Malaysia); BUAP, CINVESTAV, CONACYT, LNS, SEP, and UASLP-FAI (Mexico); MOS (Montenegro); MBIE (New Zealand); PAEC (Pakistan); MSHE and NSC (Poland); FCT (Portugal); JINR (Dubna); MON, RosAtom, RAS, RFBR, and NRC KI (Russia); MESTD (Serbia); SEIDI, CPAN, PCTI, and FEDER (Spain); MOSTR (Sri Lanka); Swiss Funding Agencies (Switzerland); MST (Taipei); ThEPCenter, IPST, STAR, and NSTDA (Thailand); TUBITAK and TAEK (Turkey); NASU (Ukraine); STFC (United Kingdom); DOE and NSF (USA).

  \hyphenation{Rachada-pisek} Individuals have received support from the Marie-Curie program and the European Research Council and Horizon 2020 Grant, contract Nos.\ 675440, 752730, and 765710 (European Union); the Leventis Foundation; the A.P.\ Sloan Foundation; the Alexander von Humboldt Foundation; the Belgian Federal Science Policy Office; the Fonds pour la Formation \`a la Recherche dans l'Industrie et dans l'Agriculture (FRIA-Belgium); the Agentschap voor Innovatie door Wetenschap en Technologie (IWT-Belgium); the F.R.S.-FNRS and FWO (Belgium) under the ``Excellence of Science -- EOS" -- be.h project n.\ 30820817; the Beijing Municipal Science \& Technology Commission, No. Z191100007219010; the Ministry of Education, Youth and Sports (MEYS) of the Czech Republic; the Deutsche Forschungsgemeinschaft (DFG) under Germany's Excellence Strategy -- EXC 2121 ``Quantum Universe" -- 390833306; the Lend\"ulet (``Momentum") Program and the J\'anos Bolyai Research Scholarship of the Hungarian Academy of Sciences, the New National Excellence Program \'UNKP, the NKFIA research grants 123842, 123959, 124845, 124850, 125105, 128713, 128786, and 129058 (Hungary); the Council of Science and Industrial Research, India; the HOMING PLUS program of the Foundation for Polish Science, cofinanced from European Union, Regional Development Fund, the Mobility Plus program of the Ministry of Science and Higher Education, the National Science Center (Poland), contracts Harmonia 2014/14/M/ST2/00428, Opus 2014/13/B/ST2/02543, 2014/15/B/ST2/03998, and 2015/19/B/ST2/02861, Sonata-bis 2012/07/E/ST2/01406; the National Priorities Research Program by Qatar National Research Fund; the Ministry of Science and Higher Education, project no. 02.a03.21.0005 (Russia); the Programa Estatal de Fomento de la Investigaci{\'o}n Cient{\'i}fica y T{\'e}cnica de Excelencia Mar\'{\i}a de Maeztu, grant MDM-2015-0509 and the Programa Severo Ochoa del Principado de Asturias; the Thalis and Aristeia programs cofinanced by EU-ESF and the Greek NSRF; the Rachadapisek Sompot Fund for Postdoctoral Fellowship, Chulalongkorn University and the Chulalongkorn Academic into Its 2nd Century Project Advancement Project (Thailand); the Kavli Foundation; the Nvidia Corporation; the SuperMicro Corporation; the Welch Foundation, contract C-1845; and the Weston Havens Foundation (USA).\end{acknowledgments}

\bibliography{auto_generated}

\cleardoublepage \appendix\section{The CMS Collaboration \label{app:collab}}\begin{sloppypar}\hyphenpenalty=5000\widowpenalty=500\clubpenalty=5000\vskip\cmsinstskip
\textbf{Yerevan Physics Institute, Yerevan, Armenia}\\*[0pt]
A.M.~Sirunyan$^{\textrm{\dag}}$, A.~Tumasyan
\vskip\cmsinstskip
\textbf{Institut f\"{u}r Hochenergiephysik, Wien, Austria}\\*[0pt]
W.~Adam, F.~Ambrogi, T.~Bergauer, M.~Dragicevic, J.~Er\"{o}, A.~Escalante~Del~Valle, R.~Fr\"{u}hwirth\cmsAuthorMark{1}, M.~Jeitler\cmsAuthorMark{1}, N.~Krammer, L.~Lechner, D.~Liko, T.~Madlener, I.~Mikulec, F.M.~Pitters, N.~Rad, J.~Schieck\cmsAuthorMark{1}, R.~Sch\"{o}fbeck, M.~Spanring, S.~Templ, W.~Waltenberger, C.-E.~Wulz\cmsAuthorMark{1}, M.~Zarucki
% ...
\vskip\cmsinstskip
\dag: Deceased\\
1:  Also at Vienna University of Technology, Vienna, Austria\\
% ...
\end{sloppypar}
%%% END EDITABLE REGION %%%
\end{document}